\documentclass[journal=jpccck,manuscript=article]{achemso}
\setkeys{acs}{articletitle=true}
\usepackage{chemformula} 
\usepackage[T1]{fontenc} 
\usepackage{amssymb}
\usepackage{amsmath}

\usepackage{multirow}
\usepackage{booktabs}
\usepackage{subcaption}

\newcommand{\RomanNumeralCaps}[1]
    {\MakeUppercase{\romannumeral #1}}
    
\author{Chao Zheng}
\affiliation[McMaster University]
{Department of Materials Science and Engineering, McMaster University, Hamilton}
\email{zhengc8@mcmaster.ca}
\author{Oleg Rubel}
\affiliation[McMaster University]
{Department of Materials Science and Engineering, McMaster University, Hamilton}
\email{rubelo@mcmaster.ca}

\title[An \textsf{achemso} demo]
  {Unraveling the Water Degradation Mechanism of \ch{CH3NH3PbI3}}


\begin{document}

\begin{tocentry}

\includegraphics{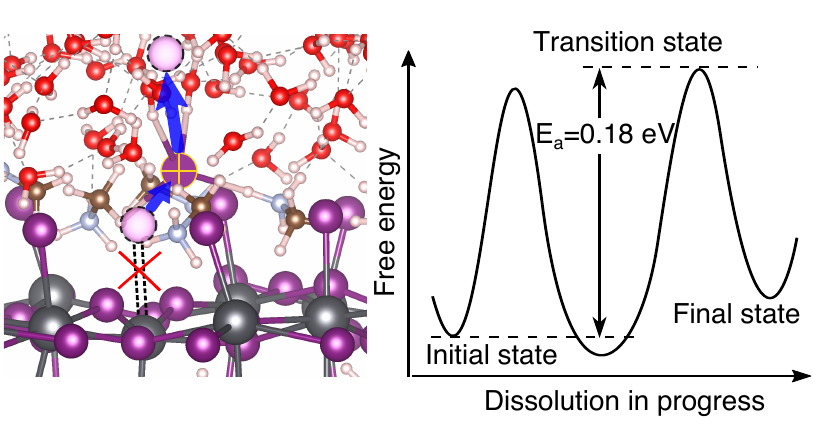}

\end{tocentry}

\begin{abstract}
Instability of perovskite photovoltaics is still a topic which is currently under intense debate, especially the role of water environment. Unraveling the mechanism of this instability is urgent to enable practical application of perovskite solar cells. Here, \textit{ab initio} metadynamics is employed to investigate the initial phase of a dissolution process of \ch{CH3NH3PbI3} (\ch{MAPbI3}) in explicit water. It is found that the initial dissolution of \ch{MAPbI3} is a complex multi-step process triggered by the departure of \ch{I-} ion from the \ch{CH3NH3I}-terminated surface. Reconstruction of the free energy landscape indicates a low energy barrier for water dissolution of \ch{MAPbI3}. In addition, we propose a two-step thermodynamic cycle for \ch{MAPbI3} dissolution in water at a finite concentration that renders a spontaneity of the dissolution process. The low energy barrier for the initial dissolution step and the spontaneous nature of \ch{MAPbI3} dissolution in water explain why the water immediately destroys pristine \ch{MAPbI3}. The dissolution thermodynamics of all-inorganic \ch{CsPbI3} perovskite is also analyzed for comparison. Hydration enthalpies and entropies of aqueous ions play an important role for the dissolution process. Our findings provide a comprehensive understanding to the current debate on water instability of \ch{MAPbI3}.
\end{abstract}

\section{Introduction}
Evolution of hybrid halide perovskite solar cells makes a contribution to the goal of replacement of fossil fuels. High power conversion efficiency and low fabrication cost raise perovskite photovoltaics as a tough competitor against the silicon solar cells. Since the inception of halide perovskite solar cells with 3.8\% power conversion efficiency in 2009\cite{Kojima_JACS_131_2009}, within 10 years of development, the world record power conversion efficiency of perovskite photovoltaics has reached 24.2\% according to the efficiency chart published by the National Renewable Energy Laboratory. However, the poor stability of perovskite photovoltaic absorbers still remains unsolved and hinders the solar cells entering people's daily life. 

Among the stimuli caused degradation of hybrid perovskite, water is confirmed to degrade \ch{MAPbI3}\cite{Hailegnaw_JPCL_6_2015}. Whereas, the role of water incorporation into \ch{MAPbI3} remains a topic of debate, with conflicting results reported in the literature. It is found the moisture invasion fractures the connection of \ch{C-N} in \ch{CH3NH3} (MA) and generate ammonia and hydrogen iodide\cite{Li_JPCC_119_2015, Ke_CC_53_2017}
\begin{equation}\label{Eq:BH_MAPbI3_decompose1}
	\text{CH}_3\text{NH}_3\text{PbI}_3(\text{s})
	\xrightarrow{\text{H}_2\text{O}}
	(-\text{CH}_2-) + \text{NH}_3(\text{g}) + \text{HI}(\text{g}) + \text{PbI}_2(\text{s})~.
\end{equation}
\citet{Kakekhani_APLM_7_2019} propose a super-hydrous state of water incorporation in \ch{MAPbI3} to explain deterioration of its photovoltaic properties in a moist environment. Besides, \citet{Philippe_CM_27_2015} and \citet{Niu_JMCA_2_2014} proposed a decomposition mechanism relating to the presence of \ch{CH3NH3I} (\ch{MAI}) and \ch{PbI2},
\begin{equation}\label{Eq:BH_MAPbI3_decompose2}
	\text{CH}_3\text{NH}_3\text{PbI}_3(\text{s})
	\xrightarrow{\text{H}_2\text{O}}
	\text{CH}_3\text{NH}_3\text{I}(\text{aq}) + \text{PbI}_2(\text{s})~.
\end{equation}

In fact, two extreme water content conditions are discussed above: atmospheric moisture and liquid water. Both of the mechanisms are important. For fresh perovskite photovoltaics, moisture degradation is obviously important for the endurance. There are plenty of publications investigating the moisture induced degradation of \ch{CH3NH3PbI3}\cite{Frost_NL_14_2014, Wei_IJER_41_2016, McLeod_JPCL_9_2018, Yang_AMI_10_2018, Kosasih_NE_47_2018, Aristidou_JMCA_5_2017, Smecca_PCCP_18_2016, Saidaminov_NE_3_2018}. As photovoltaics, it has to be considered that different situations the cells will encounter, e.g. after couple of years, cracks of the photovoltaic modules are inevitable\cite{Lin_ISOP_2015}. At this stage, understanding of liquid water degradation mechanism becomes important. Especially,  hybrid halide perovskites contain lead element. There are less investigations about the liquid water degradation pathway\cite{Mosconi_CM_27_2015, Caddeo_AMI_2018, Caddeo_AN_11_2017}. It is still unclear about the mechanism of \ch{CH3NH3PbI3} degradation in liquid water, such as the energy barrier, decomposition details.

Key challenges in understanding the degradation mechanism of \ch{MAPbI3} are the difficulties in determining the surface chemistry of the first few atomic layers of the pristine material without any exposure to ambient atmospheric conditions, and, conversely, the difficulties in measuring this same surface chemistry in real time under exposure to realistic environments\cite{Ke_CC_53_2017}. Molecular dynamics (MD) make it possible to explore the degradation mechanism at the atomic level. \citet{Mosconi_CM_27_2015, Caddeo_AN_11_2017} simulated the heterogeneous interface between water and \ch{MAPbI3} to study its water degradation mechanism. \citet{Mosconi_CM_27_2015} observed first \ch{MAI} unit dissolved within 8.5~ps using an \textit{ab initio} molecular dynamics (AIMD). \citet{Caddeo_AN_11_2017} observed a fast dissolution of outermost \ch{MAI}-terminated layers within 10~ps using classical molecular dynamics. Whereas, the details of the initial phase of the degradation process, such as energy barriers of corresponding steps, and the preference of ions leaving the surface are still missing.  We performed AIMD simulation for water and \ch{MAPbI3} interface aiming to clarify the degradation mechanism of \ch{MAPbI3} in water. Contrary to previous reports of a short degradation time, our AIMD simulation of the heterogeneous interface renders no dissolution event happening within 12~ps. This phenomenon questions the immediacy of \ch{MAPbI3} degradation in water. \cite{Hailegnaw_JPCL_6_2015} During 12~ps standard AIMD simulation, we observed both \ch{I-} and \ch{MA+} ions trying to detach from the \ch{MAPbI3} surface. As time passed by, the ions retrace back to the \ch{Pb-I} framework cavity. Thus, the dissolution event reported in Ref.~\citenum{Mosconi_CM_27_2015} could be an artefact due to a limited simulation time of about 10~ps.

The ionic nature of \ch{MAPbI3} \cite{Frost_NL_14_2014, Park_NE_2_2017} allows to draw a parallel to the research on simulation of \ch{NaCl} dissolution. Intriguingly, \citet{Liu_PCCP_13_2011} and \citet{Chen_PCCP_16_2014} encountered the same situation when dissolving \ch{NaCl} using MD. A single ionic dissolution is a rare event, which is unlikely to happen on the time scale of an AIMD simulation. Therefore, the dissolution needs to be "driven" artificially\cite{Chen_PCCP_16_2014}. Metadynamics\cite{Laio_PNAS_99_2002, Lannuzzi_PRL_90_2003, Bucko_JPCM_20_2008} is a powerful algorithm that can be used both for probing the free energy landscape and for accelerating rare events in systems described by complex Hamiltonians, at the classical or at the quantum level. Since the dissolution of \ch{MAPbI3} in water is viewed as a rare event, it needs a much longer simulation time to capture the dissolution, and this finding raises a question about validity of conclusions drawn from the relatively short 10~ps simulation\cite{Caddeo_AN_11_2017}.

In the current research, we propose the water degradation mechanism employing the capability of \textit{ab initio} metadynamics which uses computational sands to fill the initial potential valley and to force the trapped system from the initial basin and explore the energy landscape. And it gives the possibility to describe the heterogeneous interface with a large system from the atomic level, dynamically and considering finite temperature effects. Hence, the method can accelerate and capture the dissolution process of \ch{MAPbI3} in water. Here, the free energy surface (FES) of water dissolution \ch{MAPbI3} can be reconstructed based on the historical computational sands added to the basin. The obtained FES suggests a relatively low energy barrier of the first step of the dissolution process. In addition, an analysis based on a thermodynamic cycle for dissolution \ch{MAPbI3} explains the intrinsic water instability of \ch{MAPbI3}. The low dissolution energy barrier and spontaneous dissolving trend together unravel the fragile nature of \ch{MAPbI3} when encountering water.

\subsection{Computational method}
Electronic structure calculations have been performed in the framework of DFT \cite{Kohn_PR_140_1965} and Perdew-Burke-Ernzerhof generalized gradient approximation\cite{Perdew_PRL_77_1996} (GGA-PBE) for the exchange-correlation functional. Van der Waals correction is important for halide hybrid perovskite\cite{Wang_PCCP_16_2013, Zheng_9_JPCL_2018}. Among different van der Waals corrections, \citet{Li_EES_12_2019} employed both optb86B-vdW\cite{Klimevs_JPCM_22_2009} and PBE+D3 method\cite{Grimme_JCP_132_2010} to optimize structures. The results show that both the vdW functionals can give accurate crystal structure predictions. And the PBE+D3 method acts even better than optb86B-vdW when comparing with experimental data. Except for the structural properties, our previous work\cite{Zheng_9_JPCL_2018} analyzed the polymorphism of \ch{MAPbI3} employing different functionals including PBE+D3. We concluded that the PBE+D3 can accurately predict the trend of polymorphism of \ch{MAPbI3}. It indicates PBE+D3 can capture the total energy estimation very well. Hence, the PBE+D3 level  van der Waals correction is considered for all calculations. Total energies of all compounds were obtained using the Vienna \textit{ab initio} simulation program (VASP) and projector augmented-wave (PAW) potentials \cite{Kresse_PRB_54_1996,Kresse_PRB_59_1999,Blochl_PRB_50_1994}.

The phase separation energy difference $\Delta E_\text{tot}$ of \ch{MAPbI3} and \ch{CsPbI3} are adopted from our previous calculations\cite{Zheng_JPCC_131_2017}. For reciprocal space integration, $3\times3\times2$ Monkhorst-Pack grid \cite{Monkhorst_PRB_13_1976} was used for tetragonal \ch{MAPbI3}, $4\times4\times3$ for hexagonal \ch{PbI2} and $3\times6\times2$ for orthorhombic \ch{CsPbI3}. The convergence of $\Delta E_\text{tot}$ with respect to the k-mesh density was tested via doubling the density for investigated perovskite structures and corresponding decomposed structures. And the convergence is better than 5 meV. The cutoff energy for a plane wave expansion was set at 400~eV. The lattice constant and atomic positions were optimized such that residual forces acting on atoms did not exceed 2~meV/{\AA}, and the residual hydrostatic pressure was less than 50~MPa.

For AIMD calculations, a semi-empirical scaling method\cite{Zheng_PRM_2_2018, Whitfield_SR_6_2016} is used to achieve a finite-temperature structure of \ch{MAPbI3} that is self-consistent with PBE functional with the van der Waals correction. According to scanning tunnelling electron microscopy studies of \ch{MAPbI3}\cite{She_ACS_10_2015, Ohmann_JACS_137_2015}, we selected a \ch{MAI}-terminated (001) surface structure of tetragonal \ch{MAPbI3}. The \ch{MAPbI3} slab was modelled as $2 \times 2$ in plane per-optimized tetragonal supercell with the thickness of 7 atomic layers spaced by 18.6~{\AA}  filled by water molecules (see Fig.~\ref{fgr:IS}). The number of water molecules embedded is 158 which is obtained based on the experimental liquid water density. The dimension of the periodic cell are $a = b = 17.72$~{\AA} which corresponds twice of the size of tetragonal \ch{MAPbI3}. In total, we have $c = 38.35$~{\AA} for the heterogeneous structure. In order to obtain an initial randomization of the atomic positions, we performed a standard AIMD simulation in two stages: pre-heating followed by a fixed temperature relaxation. Pre-heating from 0 to 300~K was performed in 700 steps (step size of 1~fs) using a linear ramp-up function (VASP tag $\text{SMASS}=-1$). Velocities were scaled every 20~MD steps. Although the orientations of \ch{MA+} cations on surface are anisotropic after this two-stage relaxation, we noticed the \ch{-NH3+} groups of \ch{MA+} ions are attracted by oxygen atoms from ajdacent water molecules during relaxation. Accuracy of computed Hellmann-Feynman forces was determined by the energy convergence criterion of $10^{-6}$~eV. Only one $k$ point at $\Gamma$ was used to sample the Brillouin zone. Atomic positions and velocities at the end of the preheating stage were taken as the input for the fixed temperature relaxation. The fixed temperature relaxation was conducted at 300~K for $\sim{9.8}$~ps (step size of 1~fs). A Nos{\'e}--Hoover thermostat\cite{Nose_JCP_81_1984, Hoover_PRA_31_1985} was used to stabilize the temperature (VASP tag $\text{SMASS}=0$). Atomic positions during AIMD were stored every 20 steps. Crystallographic information files (CIF) with atomic structures used in calculations can be accessed through the Cambridge crystallographic data centre (CCDC deposition numbers 1919295$-$1919299).

Metadynamics was applied to accelerates the rare events of the heterogeneous interface (VASP tag $\text{MDALGO}=21$). It is realized by augmenting the system Hamiltonian $\tilde{H}(t)$  with a time-dependent bias potential $\tilde{V}(t, \xi)$ which acts on selected collective variables $\xi=\{\xi_1, \xi_2, ..., \xi_m\}$
\begin{equation}\label{Eq:Hamiltonian}
\tilde{H}(t) = H + \tilde{V}(t, \xi)~,
\end{equation}
where $H$ stands for the original Hamiltonian of unbiased system.  $\tilde{V}(t, \xi)$ is defined as a sum of Gaussian hills with height $h$ and width $w$,
\begin{equation}\label{Eq:BiasedPOT}
\tilde{V}(t, \xi) = h \sum^{\lfloor{t/t_\text{G}}\rfloor}_{i=1}
\text{exp} \left [
- \dfrac{\vert\xi^{(t)} - \xi^{(i\cdot t_G)}\vert^2}
             {2w^2}
             \right ]~.
\end{equation}
During the metadynamic simulation, $\tilde{V}(t, \xi)$ is updated by adding a new Gaussian with a time increment $t_\text{G}$ which is set to 100~fs. A collective variable (CV) is a function of the particle positions. We employed two CVs in the current metadynamics. The first CV ($\xi_1$) is defined as the coordination number
\begin{equation}\label{Eq:CV1}
\xi_1 = \sum^M_{i=1} \dfrac{1-(q_i/c_i)^9}{1-(q_i/c_i)^{14}}
\end{equation}
of the departing \ch{I-} (or N atom of monitored \ch{MA+} during a paralleled \ch{MA+} dissolution) with the rest \ch{I-} (or N atoms of other \ch{MA+}) in the topmost complete layer of the surface. $M$ is the number of \ch{I-} in the topmost layer except the monitored \ch{I-} (or N atom in \ch{MA+}). $c_i$ is defined as the  interatomic distance between the monitored \ch{I-} (or N atom in \ch{MA+}) and each rest of \ch{I-} (or N atom in \ch{MA+}) in the initial state. $q_i$ is the on-the-fly interatomic distance between the monitored \ch{I-}(or N atom in \ch{MA+}) and each rest of \ch{I-} (or N atom in \ch{MA+}) during the simulation.

The second CV ($\xi_2$) records the interatomic distance between the monitored \ch{I-} and \ch{Pb^2+} underneath it.
An estimate of the underlying free energy $A(\xi)$ can be obtained via a sufficiently long time simulation,
\begin{equation}\label{Eq:FES}
A(\xi) = \lim_{t\to\infty}  \tilde{V}(t, \xi) + \text{const}~.
\end{equation}
The choice of coordination numbers and \ch{Pb-I} distance as two CVs is because these CVs vary according to different dissolution stages. CV ($\xi_2$) directly indicates the dissolution process of the monitored \ch{I-} ion. However, only CV ($\xi_2$) is not sufficient to tell the difference between the states of \ch{I-} ion attached on surface and further dissolution in water. CV ($\xi_1$) is able to reflect the bond breaks accompanying the leaving of \ch{I-} ion. Combination with CV ($\xi_1$), it is clear to explore an intermediate state. \citet{Liu_PCCP_13_2011} used different combination of CVs and determined that the settings same as we used are the best combination to characterize the dissolution event. 

After the standard $\sim{9.8}$~ps MD, the width of Gaussian hill is determined from a continuous 1.8 ps metadynamics which monitors the two CVs without adding hills. The amplitudes of these CVs in the reactant well indicate the width of the well\cite{Ensing_PJCB_109_2005}, and we set $w = 0.11$. Considering our large system (more than 800 atoms) and complexity of the dissolution procedure, we set $h=0.026$~eV from IS to IM. After passing the IM state, we increased the Gaussian height to 0.052~eV.
To characterize the hydrogen bonds between water molecules and \ch{I-}, we set the bonding searching range to 3.25~{\AA} \cite{Karmakar_JPCB_119_2015}.
\texttt{Plumed} package\cite{Bonomi_CPC_180_2009} and \texttt{Gnuplot} were utilized to reconstruct and plot the FES of dissolution events.
\texttt{VESTA}~3 package \cite{Momma_JAC_44_2011} was used to visualize crystal structures.

\section{Results and discussion}
\subsection{Dissolution energy barrier estimation}
In this section, it will be shown that \ch{MAPbI3} dissolution is a complex multi-step process triggered by the initial departure of \ch{I-} ions from the \ch{MAI}-terminated surface. The choice of the \ch{MAI}-terminated surface as a starting point is based on the scanning tunnelling microscopy topography observations of halide hybrid perovskite surface\cite{She_ACS_10_2015, Ohmann_JACS_137_2015}. An intermediate state is identified as the departing ion is partially hydrated but still remains within a proximity from the \ch{MAPbI3} surface. Starting with an equilibrated configuration we performed metadynamics using biasing variables (see Method section) aimed at obtaining the lowest free energy pathway for the detachment of \ch{I-} from the \ch{MAPbI3} surface.

\begin{figure}[t]
  \includegraphics[width=0.5\textwidth]{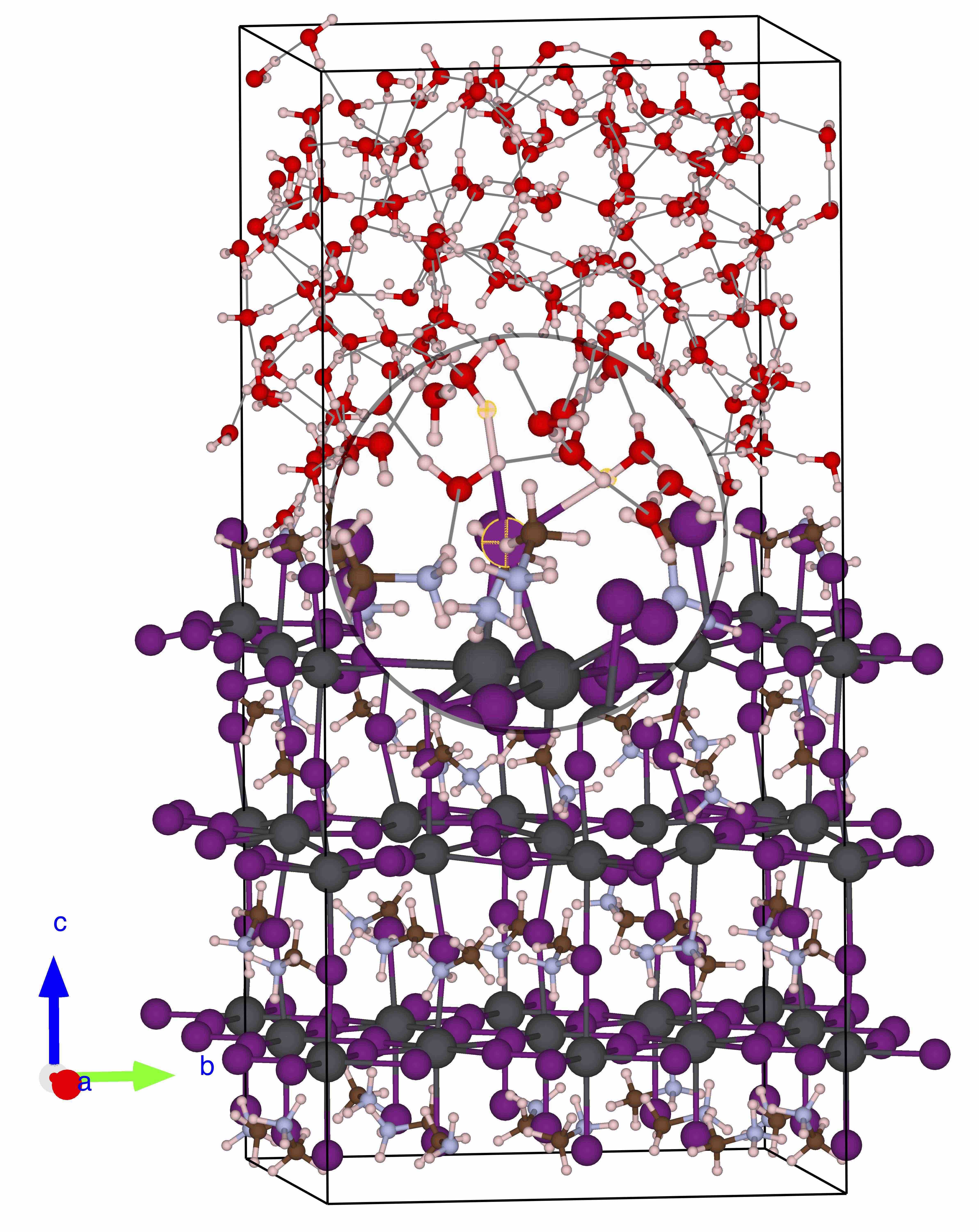}
  \caption{The initial state (IS) configuration of water/\ch{MAPbI3} interface. Light pink represents hydrogen atoms. Red represents oxygen atoms. Brown represents carbon atoms. Light purple represents nitrogen atoms. Grey represents lead atoms. Purple represents iodine atoms. }
  \label{fgr:IS}
\end{figure}

\begin{figure}[t]
  \includegraphics[width=1\textwidth]{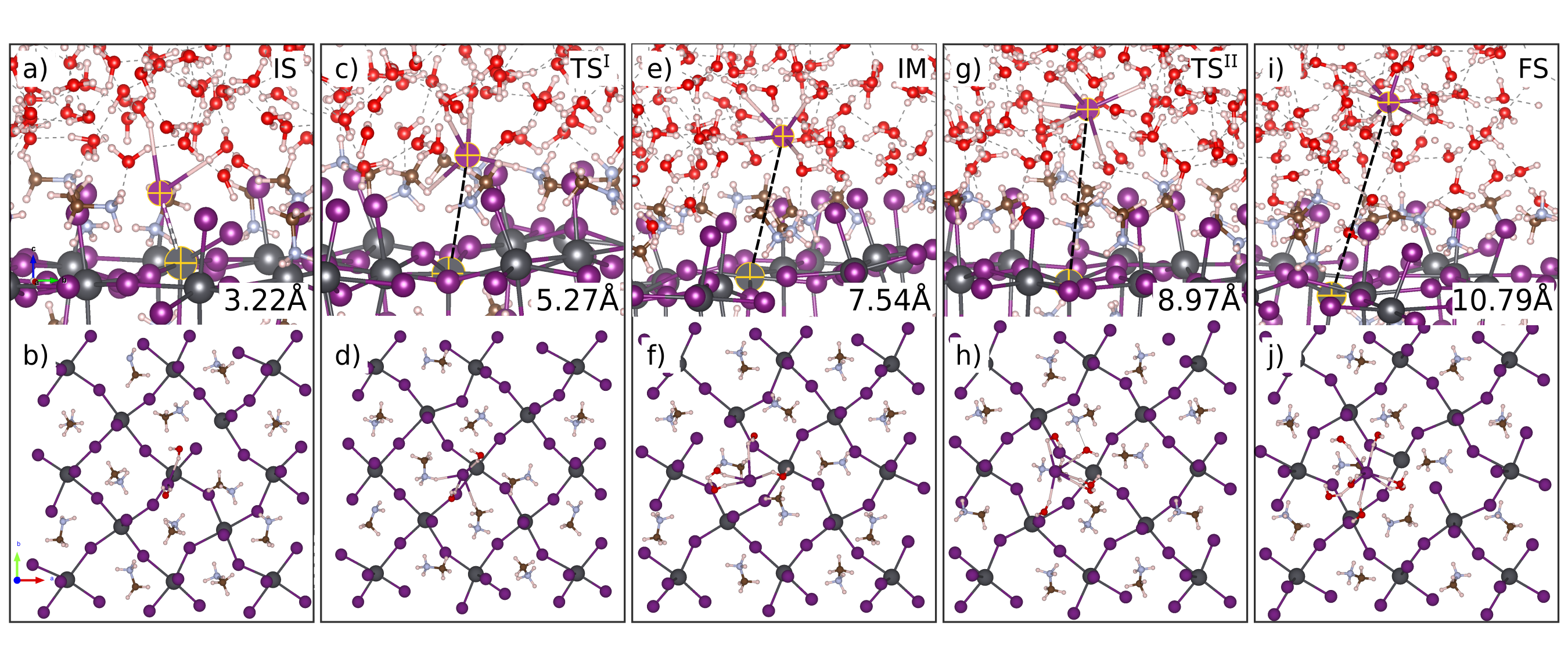}
  \caption{The initial state (IS), first transition state ($\text{TS}^{\text{\RomanNumeralCaps{1}}}$), intermediate state (IM),  second transition state ($\text{TS}^{\text{\RomanNumeralCaps{2}}}$) and final state (FS) of the initial dissolution of \ch{MAPbI3} in water, respectively. \ch{I-Pb} interaction is marked as dashed line, and the value is shown in each configuration. It is clear to see that the \ch{I-Pb} bond breaks at $\text{TS}^{\text{\RomanNumeralCaps{1}}}$ and the monitored \ch{I-} detaches surface at $\text{TS}^{\text{\RomanNumeralCaps{2}}}$. Lower panel shows the conrresponding topview of each configuration.}
  \label{fgr:DissolutionConf}
\end{figure}

We begin with the discussion of dissolving \ch{I-}  using the metadynamics. An equilibrated heterogeneous interface is taken as the initial structure for metadynamic simulation of the dissolution process. The observation of \ch{I-} or \ch{MA+} ion backtracking to \ch{Pb-I} cavity in the previous discuss is due to the trapping of the system in the initial FES basin using standard AIMD. The configuration trapped at this basin is named as the initial state (IS) shown in Fig.~\ref{fgr:IS}.
In the IS, the monitored \ch{I-} (labelled yellow) forms hydrogen bonds with two water molecules above it, and bonds with a lead atom underneath it. Under the action of the metadynamic bias, the system is discouraged to revisit previous explored spots. With accumulating the computational sands, the initial basin is filled and the system is forced to escape from the local minima. We observe the elongation of the \ch{I-Pb} bond shown in Fig.\ref{fgr:DissolutionConf}a and Fig.\ref{fgr:DissolutionConf}b. After around 7~ps of metadynamic simulation, the system arrives at the first transition state ($\text{TS}^{\text{\RomanNumeralCaps{1}}}$) of the dissolution process shown in Fig.~\ref{fgr:DissolutionConf}c and Fig.~\ref{fgr:DissolutionConf}d, the \ch{I-Pb} bond breaks.
At the transition state, the \ch{I-} ion retains two hydrogen bonds with two water molecules. Besides, the \ch{I-} forms bonds with a hydrogen atom on \ch{-NH3+} group of one \ch{MA+}, and with a hydrogen atom on \ch{-CH3} group of the another \ch{MA+}. The interatomic distance of the monitored \ch{I-Pb} bond streches from 3.22~{\AA} at IS to 5.27~{\AA} at $\text{TS}^{\text{\RomanNumeralCaps{1}}}$. We continued the metadynamic simulation after conquering the $\text{TS}^{\text{\RomanNumeralCaps{1}}}$. The lifted \ch{I-} drifts away from the underneath \ch{Pb^2+}. Meanwhile, neighbour \ch{MA+} cations of the monitored \ch{I-} drift towards the cavity. It is intriguing to find that the departing \ch{I-} ion does not enter solvent immediately and the surrounding \ch{MA+} cations do not leave the lattice following the \ch{I-} ion. The heterogeneous interface evolves towards a state in which the \ch{I-} is partly hydrated, but remains trapped close to the surface illustrated in Fig.~\ref{fgr:DissolutionConf}e and Fig.~\ref{fgr:DissolutionConf}f. We assign this local energy minimum as an intermediate state (IM). The partially hydrated \ch{I-} is in an adatom-like configuration. Upon transition from $\text{TS}^{\text{\RomanNumeralCaps{1}}}$ to IM, the coordination number of the \ch{I-} with the solvent water molecules increases from $\sim{2}$ to $\sim{4}$.

During the whole simulation, we find the system spends a long time at IM basin. The partly hydrated \ch{I-} stays on top of one neighbour \ch{MA+} cation due to an electrostatic attraction. Once the IM basin is filled, the system comes to the second transition state ($\text{TS}^{\text{\RomanNumeralCaps{2}}}$), the corresponding configuration is shown in Fig.~\ref{fgr:DissolutionConf}g and Fig.~\ref{fgr:DissolutionConf}h. From IM to $\text{TS}^{\text{\RomanNumeralCaps{2}}}$, the pulls from water molecules acting on the \ch{I-} gradually overwhelm the interactions between the \ch{I-} and underneath \ch{MA+}. The adatom-like \ch{I-} ion detaches from the \ch{MA+} cation. The number of water molecules in the hydration shell of \ch{I-} increase to 5. The system needs to conquer the barrier of 0.22~eV to reach $\text{TS}^{\text{\RomanNumeralCaps{2}}}$. This relatively larger energy barrier (compared with the initial 0.16~eV) is explained by breaking of the electrostatic attraction between the leaving \ch{I-} and \ch{MA+}. After passing through the $\text{TS}^{\text{\RomanNumeralCaps{2}}}$, the system evolves to a final state (FS) shown in Fig.~\ref{fgr:DissolutionConf}i and Fig.~\ref{fgr:DissolutionConf}j. We characterize the FS as a state where \ch{I-} escapes from the \ch{MAPbI3} surface and fully dissolves in water. In the FS, \ch{I-} ion is coordinated by $\sim{7}$ water molecules, consistent with both experimental and \textit{ab initio} simulated coordination number of  $6-9$\cite{Markovich_JCP_95_1991, Karmakar_JPCB_119_2015}. The overall energy barrier for the initial dissolution event is obtained as the energy difference between the IS and $\text{TS}^{\text{\RomanNumeralCaps{2}}}$ which is 0.18~eV.

\begin{figure}[t]
  \includegraphics[width=1\textwidth]{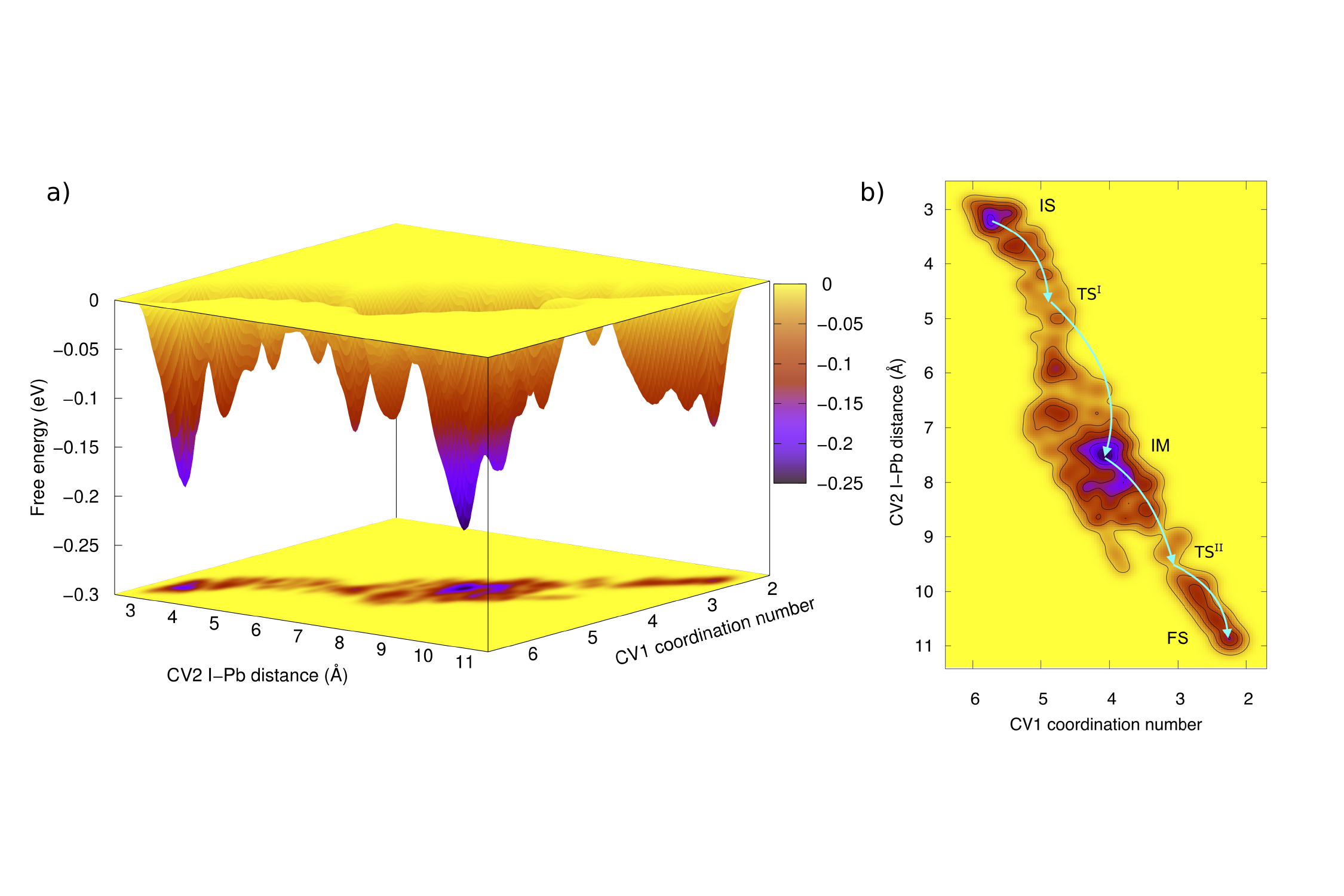}
  \caption{Reconstructed free energy surface of dissolution of \ch{MAPbI3} in water from metadynamic simulation. a) gives the free energy landscape of the dissolution process of \ch{MAPbI3} in water. b) is the contour plot of the free energy surface. Arrows point steps on the dissolution pathway.}
  \label{fgr:FES}
\end{figure}

We portray the FES of the dissolution process of \ch{MAPbI3} in water in Fig.~\ref{fgr:FES}.
The ease of the first dissolution event happening is determined by an energy barrier between IS and $\text{TS}^{\text{\RomanNumeralCaps{1}}}$. Figure~\ref{fgr:FES} suggests that the initial configuration needs to conquer 0.16~eV energy barrier to break the \ch{I-Pb} bond in aqueous environment to reach $\text{TS}^{\text{\RomanNumeralCaps{1}}}$. For comparison, \citet{Caddeo_AN_11_2017} reported an energy barrier of 0.36~eV consisting of a layer-by-layer degradation of \ch{MAPbI3} in water. \citet{Lin_NM_17_2018} reported a larger energy barrier (greater than 0.5~eV) when dissolving an \ch{I-} from \ch{CsPbI3} surface in water. Releasing the \ch{I-} is a competition between the hydration by water molecules and the electrostatic attraction to atoms on the surface and in the bulk of \ch{MAPbI3}. From a thermodynamic point of view, the dissolution phenomenon is an effect of compensating electrostatic energy stored in the \ch{MAPbI3} surface by heat released from the hydration of the \ch{I-}. This process is further promoted by the entropy gain during \ch{I-} entering water. Detailed thermodynamic analysis of the overall dissolution process is discussed in the following section. Both the hydrophilicity of \ch{I-} and \ch{MA+} ion and the intrinsic low electrostatic energy of \ch{MAPbI3}\cite{Frost_NL_14_2014, Zheng_JPCC_131_2017} contribute to these very low energy barriers in Fig.~\ref{fgr:FES}. 

Simulations of \ch{NaCl} dissolution in water identify a clear preference for the dissolution of \ch{Cl-} as the initial step over \ch{Na+}\cite{Ohtaki_PAC_60_1988, Yang_PRE_72_2005, Liu_PCCP_13_2011}. To investigate the preference of ions leaving \ch{MAPbI3} surface at the beginning of the dissolution process, we performed an alternative metadynamic simulation where the dissolution starts with \ch{MA+}. The FES of dissolution of \ch{MA+} is constructed and it renders a deeper energy basin ($\sim{0.31}$~eV), shown in Fig. S1. However, we only obtained the FES for the inital basin of the dissolution of \ch{MA+}. We find that the initial basin of \ch{MA+} dissolution is much wider and deeper than the case of \ch{I-} ion. Although we use relative large Gaussian height and weight, it is hard to explore the sequent stages of \ch{MA+} dissolution in water. A larger Gaussian will accelerate the exploration, however, sacrifice of accuracy is expected. Due to the capability of our facility and limited time, we stopped here for exploring the sequent FES of MA+ dissolution in water. The shape and rotational property of \ch{MA+} ion may contribute to the large initial basin.

The difference of the initial basins indicates \ch{I-} is relatively easier to dissolve in water. Although \ch{-NH3+} group of \ch{MA+} is hydrophilic, the connected \ch{-CH3} group is hydrophobic. When dissolving \ch{MA+} in water, the hydrophobic \ch{-CH3} group needs more space to settle in water molecules\cite{Fedotova_RCB_61_2012}. Seeking for more space in water molecule network and breaking the interactions of \ch{MA+} with surrounding ions in \ch{MAPbI3} surface together lead to a higher energy barrier for the first dissolution of \ch{MA+}. As a result, the lower energy barrier suggests a priority of \ch{I-} leaving the surface. Note that in the supporting information of \citet{Caddeo_AN_11_2017}, their simulation indicates that the first step of the dissolution \ch{MAI}-terminated surface in water is releasing \ch{MA+} at 5~ps. Differently, during our over 33.5~ps simulation time including equilibrium relaxation and metadynamics, we only observe the dissolution of one \ch{I-} ion in water. \citet{Liu_PCCP_13_2011} indicated a forcefield-based description of \ch{NaCl} dissolution in water failed to capture a preference for \ch{Cl-} over \ch{Na+} dissolution. Limitations related to a proper description of bond breaking or formation in the empirical potential framework may contribute to this discrepancy.

It is worthy to note that we didn't observe any decomposition of \ch{MA+} cation during the whole simulation. Compared with the high deprotonation energy ($\sim$4.03~eV) of \ch{MA+} cation\cite{Delugas_PRB_92_2015}, the low energy barrier of releasing \ch{I-} suggests the initial degradation \ch{MAPbI3} in water is the dissolution of  \ch{MAI}-terminated layer into water solute. It is meaningful to mention that the energy barrier of transformation of the \ch{PbI2} 2D-planer layer to \ch{PbI2} trigonal configuration is around 0.26~eV \cite{Fan_JOULE_1_2017}. Considering the low energy barriers of dissolving \ch{I-} in water as well the low energy barriers of decomposition of \ch{Pb-I} layer, the overall degradation energy barrier of \ch{MAPbI3} dissolution in water is less than 0.3~eV.

 \subsection{Thermodynamics of \ch{MAPbI3} dissolution in water}
Followed by the discussion of the initial process of \ch{MAPbI3} dissolution in water, the thermodynamic analysis of the overall decomposition is proceeded in this section. It is also intriguing to investigate how \ch{CsPbI3} reacts with water as a comparison. The dissolution of \ch{MAPbI3} via water can be described as follow:
 \begin{equation}\label{Eq:HC_MAPbI3_step0}
    \text{CH}_3\text{NH}_3\text{PbI}_3(\text{s})	
	\xrightarrow{\Delta G_\text{diss}}
	\text{CH}_3\text{NH}_3^+(\text{aq}) + \text{I}^-(\text{aq}) + \text{PbI}_2(\text{s})~.
\end{equation}

Estimation of the change of the Gibbs free energy($\Delta G^\circ_T$) between the reactants and products is a standard approach for predicting whether a  reaction or process will occur spontaneously. Combination of the density functional theory (DFT) calculation with additional thermodynamic data (see Chap.~7 in Ref.~\citenum{Sholl_2011}) is used to predict $\Delta G^\circ_T$ of phase separation of \ch{MAPbI3}\cite{Tenuta_SR_6_2016}. \citet{Kye_JPCC_2019} utilize an \textit{ab initio} thermodynamic formalism with the effect of solution to investigate the behaviour of defects on phase stability of \ch{CsPbI3}. Here, we apply the combination of theoretical and experimental data to estimate the Gibbs free energy change ($\Delta G_\text{diss}$) of \ch{MAPbI3} dissolution in water at temperature $T$ as stated in Eq.~(\ref{Eq:HC_MAPbI3_step0}). According to \citet{Sholl_2011}, $G^\circ_T$ can be expressed as,
\begin{equation}\label{Eq:Gibbs_free_energy}
	G^\circ_T = E_\text{tot} + \tilde{\mu}^\circ_{T}~.
\end{equation}
Here, $E_\text{tot}$ is the standard state enthalpy at zero temperature which is evaluated based on DFT total energy calculations. $\tilde{\mu}^\circ_{T}$ captures finite temperature effects on the chemical potentials of species involved which is evaluated from NIST-JANAF thermochemical tables as well as other experimental resources.

\begin{figure}[t]
  \includegraphics[width=0.6\textwidth]{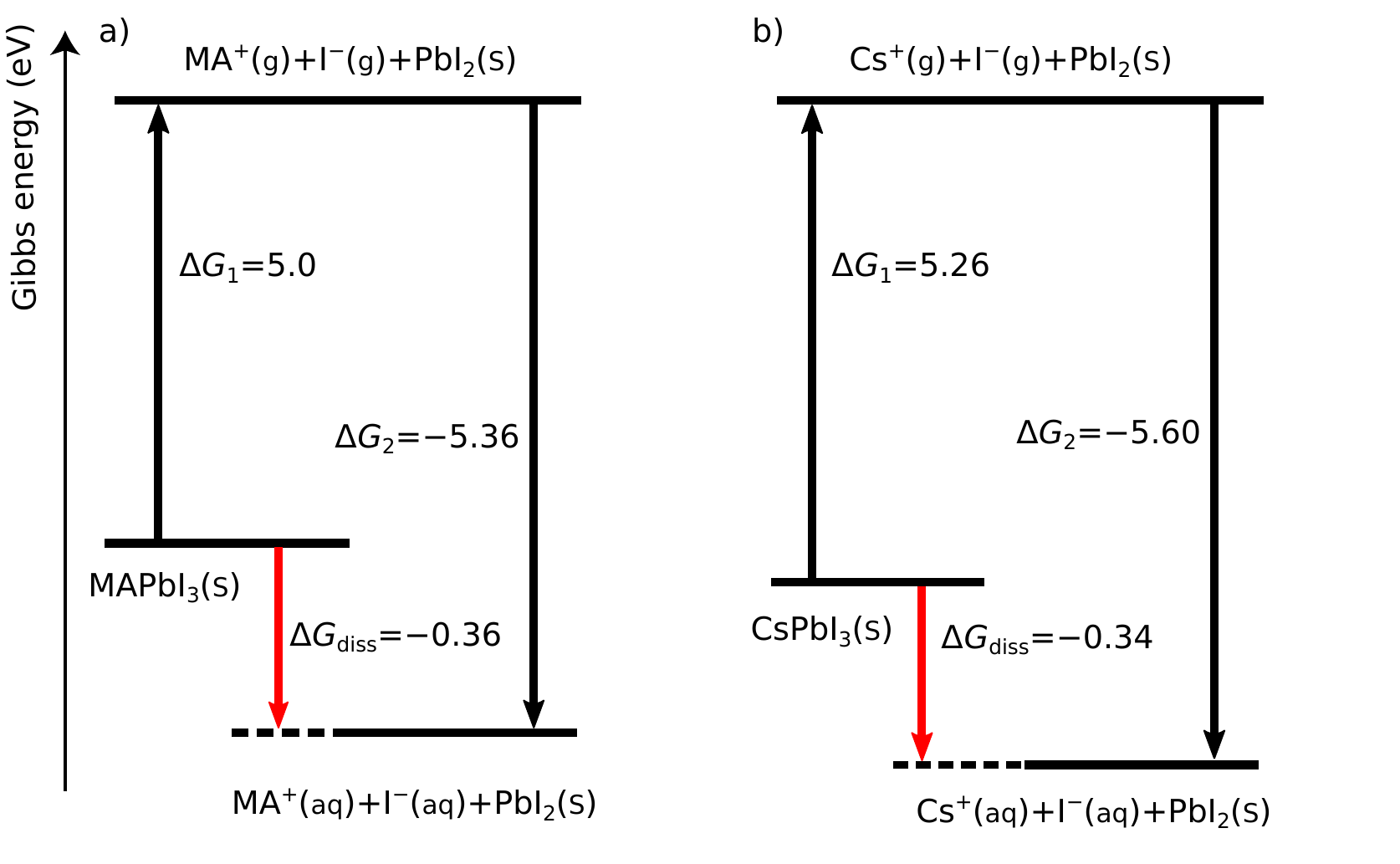}
  \caption{A thermodynamic cycle for the calculation of the dissolution Gibbs free energy change ($\Delta G_\text{diss}$) for (a) \ch{MAPbI3} and (b) \ch{CsPbI3} in water.}
  \label{fgr:Dissolution_cycle}
\end{figure}

In order to capture $\Delta G_\text{diss}$ at finite temperature, we designed a two-step thermodynamic cycle as shown in Fig.~\ref{fgr:Dissolution_cycle}a.
The first step of the thermodynamic cycle is the dissociation of \ch{MAPbI3}(s) to oppositely charged ions \ch{MA+}(g) and \ch{I-}(g) and \ch{PbI2}(s),
 \begin{equation}\label{Eq:HC_MAPbI3_step1}
    \text{CH}_3\text{NH}_3\text{PbI}_3(\text{s})	
	\xrightarrow{\Delta G_1}
	\text{CH}_3\text{NH}_3^+(\text{g}) + \text{I}^-(\text{g}) + \text{PbI}_2(\text{s})~.
\end{equation}
During this process, finite temperature effect on enthalpies and entropies of the system from solid inital state to intermediate gaseous state and \ch{PbI2}(s) are evaluated, repectively
\begin{equation}\label{Eq:chemical-potential-finite-temp-correction}
	\tilde{\mu}^\circ_{T} = H^\circ_T - H^\circ_\text{0~K} - TS^\circ_T~.
\end{equation}
The addition of the resultant $\tilde{\mu}^\circ_{T}$ and calculated $E_\text{tot}$ generates $G^\circ_T$ of each species. Related thermodynamic properties of species involved in the first step of the thermodynamic cycle of \ch{MAPbI3} dissolution in water are listed in Table~\ref{Table:step1_MAPbI3}.
\begin{table*}
    \caption{Electronic total energy $E_\text{tot}$ per formula unit (f.u.), $H_T - H_\text{0K}$, standard state entropy $S^\circ_T$, the chemical potential $\tilde{\mu}^\circ_T$ and standard Gibbs free energy $G^\circ_T$ in the first step of the thermodynamic cycle involving \ch{MAPbI3} dissolution in water. Here, we focus on room temperature $T=298.15$~K.}\label{Table:step1_MAPbI3}
        \begin{tabular}{r c c c c c c c}
        \hline
            Species & $E_\text{tot}$  & $H_T - H_\text{0K}$ & $S^\circ_T$ & $\tilde{\mu}^\circ_T$ & $G^\circ_T$ \\
            & (eV/f.u.) & (kJ/mol) & (J~$\text{mol}^{-1}~\text{K}^{-1})$ & (eV/f.u.) & (eV/f.u.)  \\
            \hline
             \ch{MAPbI3} (s)         & $-$50.93 & 44.79\textsuperscript{\emph{a}} & 374.13\textsuperscript{\emph{a}} &  $-$0.69 & $-$51.62	\\
            \ch{PbI2} (s)              & $-$8.63 & 19.50\textsuperscript{\emph{b}} & 174.84\textsuperscript{\emph{b}} & $-$0.34 & $-$8.97 \\
            \ch{MA+} (g)              & $-$32.84 & 6.20\textsuperscript{\emph{b}} & 327.7\textsuperscript{\emph{b}} &  $-$0.95 &  $-$33.79 \\
            \ch{I-} (g)              & $-$3.40 & 6.20\textsuperscript{\emph{b}} & 169.26\textsuperscript{\emph{b}} &  $-$0.46 &  $-$3.86 \\
            \hline
             \ch{CsPbI3} (s)         & $-$14.25 &20.28\textsuperscript{\emph{c}} & 219.61\textsuperscript{\emph{c}} &  $-$0.47\textsuperscript{\emph{c}} & $-$14.72\\
             \ch{Cs+} (g)              & 3.83 & 6.20\textsuperscript{\emph{b}} & 169.84\textsuperscript{\emph{b}} &  $-$0.46 &  3.37 \\
            \hline
            \ch{MAI}(s)           & $-$42.36 & 22.25\textsuperscript{\emph{d}} & 159.7\textsuperscript{\emph{d}} & $-$0.26 & $-$5.69\\
            \ch{CsI}(s)           & $-$5.45 & 13.50\textsuperscript{\emph{e}} & 123.1\textsuperscript{\emph{e}} & $-$0.24 & $-$42.62
        \end{tabular}

        \textsuperscript{\emph{a}} Data obtained from Ref.~\citenum{Onoda_JPCS_51_1990};
        \textsuperscript{\emph{b}} Data extracted from the NIST-JANAF thermochemical tables;
        \textsuperscript{\emph{c}} Data extracted from Refs.~\citenum{Ong_CMS_97_2015, Hinuma_CMS_128_2017};
        \textsuperscript{\emph{d}} Data obtained from Ref.~\citenum{Yamamuro_JCT_18_1986};
        \textsuperscript{\emph{e}} Data obtained from Ref.~\citenum{Rumble_CRC_2017}.
\end{table*}
According to Table~\ref{Table:step1_MAPbI3}, it is feasible to calculate the change in Gibbs free energy $\Delta G_1$  for the first step 
\begin{equation}\label{Eq:G_MAPbI3_step1}
	\Delta G_1 =  G^\circ_{T,\text{MA}^+(\text{g})} +  G^\circ_{T,\text{I}^-(\text{g})}+ G^\circ_{T,\text{PbI}_2} -G^\circ_{T,\text{MAPbI}_3}~.
\end{equation}
The calculated $\Delta G_1 = 5.0$~eV is shown in Fig.~\ref{fgr:Dissolution_cycle}a. The second step of the designed cycle of \ch{MAPbI3} dissolution in water involves the hydration of \ch{MA+}(g) and \ch{I-}(g)
\begin{equation}\label{Eq:HC_MAPbI3_step2}
	\text{MA}^+(\text{g}) + \text{I}^-(\text{g}) + \text{PbI}_2(\text{s})	
	\xrightarrow{\Delta G_2}
	\text{MA}^+(\text{aq}) + \text{I}^-(\text{aq}) + \text{PbI}_2(\text{s})~.
\end{equation}
The chemical potentials of aqueous ions $\tilde{\mu}^\circ_T$(aq) in this step can be obtained via
\begin{equation}\label{Eq:chemical-potential-finite-temp-correction2}
	\tilde{\mu}^\circ_{T}(\text{aq}) = \tilde{\mu}^\circ_{T} + \Delta H_\text{hyd} - T(S^\circ_T(\text{aq}) - S^\circ_T)~.
\end{equation}
$S^\circ_T$(aq) is the entropy of an aqueous ion. Decreased enthalpies and entropies due to hydration process contribute to the values of $G^\circ_{T,\text{MA}^+(\text{aq})}$ and $G^\circ_{T,\text{I}^-(\text{aq})}$. Related thermodyanmic properties of the second step are shown in Table~\ref{Table:step2_MAPbI3}.

\begin{table*}
    \caption{Hydration enthalpy $\Delta H^\circ_\text{hyd}$, entropy $S^\circ_T$(aq) of aqueous ions, and Gibbs free energy $G^\circ_T$ of species involved in the second step of the thermodynamic cycle of \ch{MAPbI3} dissolution in water at room temperature 298.15~K.}\label{Table:step2_MAPbI3}
        \begin{tabular}{r c c c c c}
        \hline
            Species & $\Delta H^\circ_\text{hyd}$ & $S^\circ_T$(aq) & $\tilde{\mu}^\circ_T$(aq)  & $G^\circ_T$\\
            & (kJ/mol) & (J~$\text{mol}^{-1}~\text{K}^{-1})$ & (eV/f.u.) & (eV/f.u.) \\
            \hline
            \ch{MA+} (aq)              & $-$284.6\textsuperscript{\emph{a}} & 142.7\textsuperscript{\emph{b}} & $-$3.33 & $-$36.17 \\
            \ch{I-} (aq)              & $-$305.0\textsuperscript{\emph{c}} & 111.3\textsuperscript{\emph{d}} & $-$3.44 & $-$6.84  \\
            \hline
            \ch{Cs+} (aq)              & $-$264.0\textsuperscript{\emph{c}} & 133.1\textsuperscript{\emph{d}} & $-$3.08 & 0.75  \\
            \hline
        \end{tabular}

        \textsuperscript{\emph{a}} Data obtained from Ref.~\citenum{Housecroft_RSCA_7_2017};
        \textsuperscript{\emph{b}} Data extracted from Ref.~\citenum{Marcus_APRC_81_1984};
        \textsuperscript{\emph{c}} Data extracted from Ref.~\citenum{Smith_JCE_54_1977};
        \textsuperscript{\emph{d}} Data extracted from Ref.~\citenum{Rumble_CRC_2017}.
\end{table*}

The change of the Gibbs free energy of the second step
\begin{equation}\label{Eq:G_MAPbI3_step2}
	\Delta G{_2} =  G^\circ_{T,\text{MA}^+(\text{aq})} +  G^\circ_{T,\text{I}^-(\text{aq})} - G^\circ_{T,\text{MA}^+(\text{g})} +  G^\circ_{T,\text{I}^-(\text{g})}
\end{equation}
amounts to $\Delta G_2 = -5.36$~eV. The strongly negative change of Gibbs free energy in the second step overcomes the Gibbs free energy gain in the first step. Combining the two-step Gibbs free energy change, we can obtain $\Delta G_\text{diss} = \Delta G{_1} + \Delta G{_2} = -0.36$~eV. The negative $\Delta G_\text{diss}$ of the dissolution \ch{MAPbI3} in water at a finite concentration indicates the reaction in Eq.~(\ref{Eq:HC_MAPbI3_step0}) would proceed spontaneously. The thermodynamic analysis of \ch{MAPbI3} dissolution process suggests an intrinsic water instability of \ch{MAPbI3}. Although the negative $\Delta G_\text{diss}$  reflects a spontaneity of a reaction, it only predicts the trend of the proposed reaction. According to Arrhenius equation, the rate of a reaction is controlled by the energy barrier, i.e activation energy. A low activation energy indicates a high rate constant. The low activation energy obtained from metadynamic calculations for the \ch{MAPbI3} degradation in water demonstrates the corresponding reaction will proceed quickly. In all, for the reaction of \ch{MAPbI3} in water, the negative $\Delta G_\text{diss}$ renders a thermodynamic instability and the low energy barrier points to a kinetic instability.

Compared with a large amount of discussions on the fragility of \ch{MAPbI3} in water, stability of \ch{CsPbI3} in water recently attracts attention and is also under discussion. \citet{Lin_NM_17_2018} indicated the water invasion triggered the phase transition of \ch{CsPbI3} from a high-temperature cubic phase to a low-temperature orthorhombic phase and they found that water is adsorbed on the surface without penetrating the interior of \ch{CsPbI3}. Conversely, \citet{Yuan_JPCC_122_2017} observed the \ch{CsPbI3} quantum dots degraded in a chamber with a wet gas flow. They confirmed the moisture was responsible for the degradation of these \ch{CsPbI3} quantum dots. Here, we employ the two-step thermodynamic cycle to the case of \ch{CsPbI3} dissolution in water to clarify these controversies.

The first step of the proposed thermodynamic cycle of dissolution \ch{CsPbI3} in water is given as
\begin{equation}\label{Eq:HC_CsPbI3_step1}
    \text{Cs}\text{PbI}_3(\text{s})	
	\xrightarrow{\Delta G_1}
	\text{Cs}^+(\text{g}) + \text{I}^-(\text{g}) + \text{PbI}_2(\text{s})~.
\end{equation}
Related thermodynamic properties of species involved in the first step of \ch{CsPbI3} dissolution in water  are listed in Table~\ref{Table:step1_MAPbI3}. We can obtain the Gibbs free energy change $\Delta G_1 = 5.26$~eV for Eq.~(\ref{Eq:HC_CsPbI3_step1}). The thermodynamic cycle of \ch{CsPbI3} is shown in Fig.~\ref{fgr:Dissolution_cycle}b.
The second step of \ch{CsPbI3} dissolution in water involves the hydration of \ch{Cs+}(g) and \ch{I-}(g)
\begin{equation}\label{Eq:HC_CsPbI3_step2}
	\text{Cs}^+(\text{g}) + \text{I}^-(\text{g}) + \text{PbI}_2(\text{s})	
	\xrightarrow{\Delta G_2}
	\text{Cs}^+(\text{aq}) + \text{I}^-(\text{aq}) + \text{PbI}_2(\text{s})~.
\end{equation}
Related thermodyanmic properties are shown in Table~\ref{Table:step2_MAPbI3}. Using Gibbs free energies in Table~\ref{Table:step1_MAPbI3} and Table~\ref{Table:step2_MAPbI3}, the change of the Gibbs free energy of the second step is estimated as $\Delta G{_2} = -5.60$~eV. The overall $\Delta G_\text{diss} = \Delta G{_1} + \Delta G{_2} = -0.34$~eV for the dissolution \ch{CsPbI3} in water
\begin{equation}\label{Eq:HC_CsPbI3_step0}
	\text{Cs}\text{PbI}_3(\text{s})	
	\xrightarrow{\Delta G_\text{diss}}
	\text{Cs}^+(\text{aq}) + \text{I}^-(\text{aq}) + \text{PbI}_2(\text{s})~.
\end{equation}

The above discussion indicates that the orthorhombic \ch{CsPbI3} is also prone to decompose in water. The thermodynamic analysis of \ch{CsPbI3} dissolution in water corroborates the degradation of \ch{CsPbI3} quantum dots observed by \citet{Yuan_JPCC_122_2017}. The energy barrier of the dissolution process determines the rate of the reaction. Compared with the very low energy barrier (about 0.18~eV) for the initial dissolution of \ch{MAPbI3}, the relatively high energy barrier (greater than 0.5~eV\cite{Lin_NM_17_2018}) for releasing \ch{I-} into water from \ch{CsPbI3} surface suggests a slow process of dissolution of \ch{CsPbI3} which explains the differences  in degradation rates at heterogeneous interfaces: water/\ch{MAPbI3} \textit{vs} water/\ch{CsPbI3} observed by \citet{Lin_NM_17_2018}.

\begin{figure}[t]
  \includegraphics[width=0.7\textwidth]{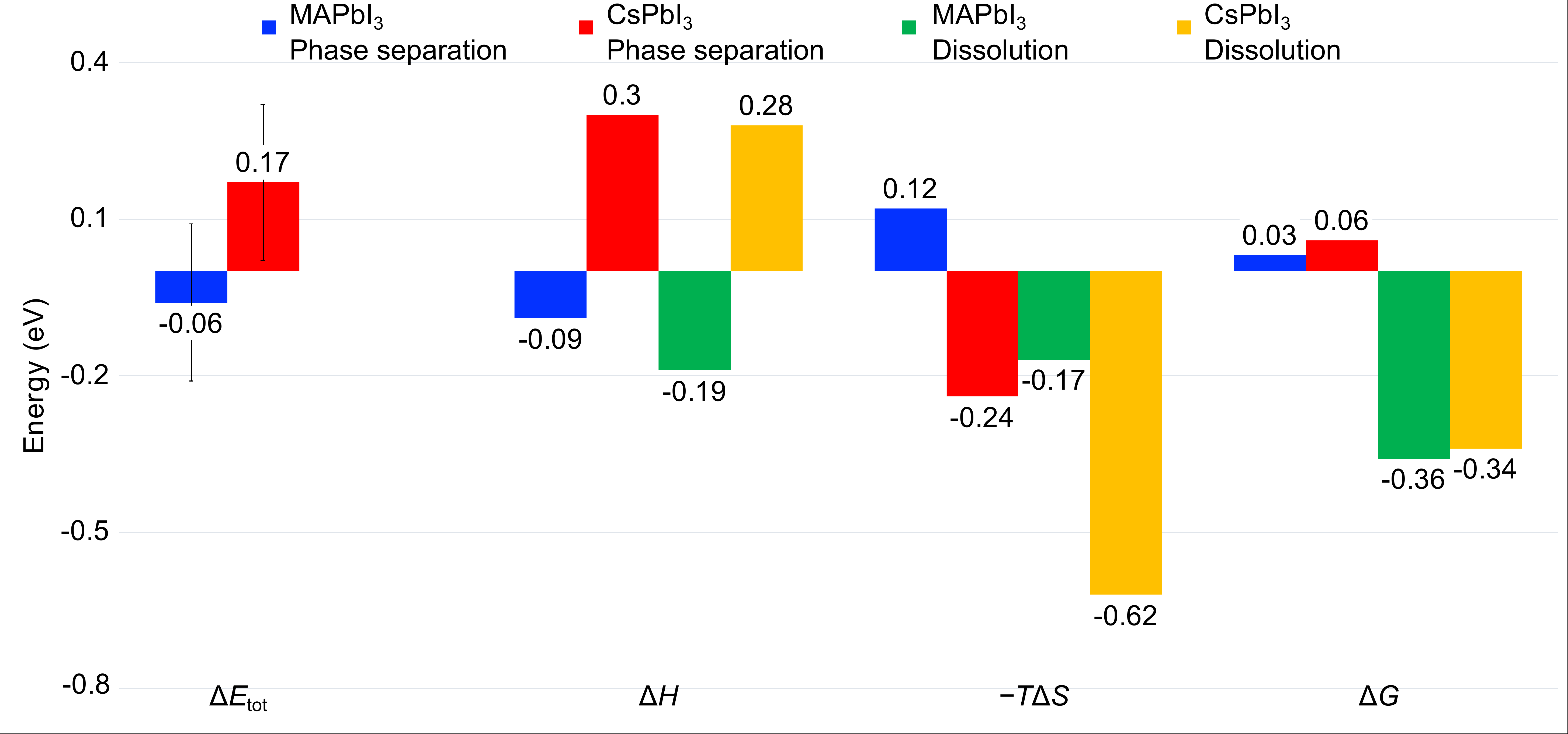}
  \caption{Thermodynamic quantities for decomposition of \ch{MAPbI3} and \ch{CsPbI3} via two alternative routes: a phase separation [Eq.~(\ref{Eq:Decomosition_reaction_general})] or a dissolution in water [Eq.~(\ref{Eq:HC_MAPbI3_step0})]. $\Delta E_\text{tot}$ is the bare-DFT energy difference between products and the reactant (perovskite). The errorbars indicate a chemical uncertainty of the DFT exchange-correlation functional (see text for details). $\Delta H$ is the enthalpy change at $T=298.15$~K. $T\Delta S$ captures the entropy change during decomposition. $\Delta G$ is the resultant Gibbs free energy change of \ch{MAPbI3} and \ch{CsPbI3} phase separation and dissolution in water.}
  \label{fgr:EC_MAPI_CSPI}
\end{figure}

\citet{Zhang_CPL_35_2018} first used the energy differences obtained from DFT calculation to characterize the intrinsic instability of \ch{MAPbI3} and \ch{CsPbI3}, considering a decomposition reaction of \ch{MAPbI3} into solid state products
\begin{equation}\label{Eq:Decomosition_reaction_general}
	\ch{MAPbI_3}(\text{s}) \rightarrow  \ch{MAI}(\text{s}) + \ch{PbI2}(\text{s}).
\end{equation}
In Fig.~\ref{fgr:EC_MAPI_CSPI} we show our calculated $\Delta E_\text{tot}$ for \ch{MAPbI3} and \ch{CsPbI3} which match well with $\Delta E_\text{tot}$ from \citet{Zhang_CPL_35_2018}. According to \citet{Zhang_CPL_35_2018}, a positive value of $\Delta E_\text{tot}$ corresponds to a stable perovskite structure.
They proposed the entropy term of \ch{MAI} is higher than that of \ch{MAPbI3} which will further destabilize \ch{MAPbI3}. However, from the measurements of entropies of \ch{MAI} and \ch{MAPbI3} listed in Table~\ref{Table:step1_MAPbI3}, it clearly indicates a reverse trend. To note that, the entropy contribution of \ch{PbI2} cannot be neglected. The overall entropy in reaction (\ref{Eq:Decomosition_reaction_general}) decreases and stabilizes \ch{MAPbI3} which has been indicated in Ref.~\citenum{Tenuta_SR_6_2016}. The entropy gains during \ch{CsPbI3} phase separation which is contrary to \ch{MAPbI3}. The enthalpy change of \ch{CsPbI3} phase separation is large enough to overcome this entropy gain resulting in a positive $\Delta G$ that is slightly greater than the value of \ch{MAPbI3}.

The positive $\Delta G$ values for phase separation process in Fig.~\ref{fgr:EC_MAPI_CSPI} indicate a feasibility of synthesis of \ch{MAPbI3} and \ch{CsPbI3} compounds. However, the result should be taken with caution, since the uncertainty in reaction energies obtained with Perdew, Burke, and Ernzerhof (PBE) \cite{Perdew_PRL_77_1996} exchange-correlation functional is of the order of $\pm0.03$~eV/atom~\cite{Hautier_PRB_85_2012}. With five atoms per perovskite formula unit (\ch{MA+} is considered as one cation), the estimated error is about $\pm0.15$~eV as shown by the errorbars in Fig.~\ref{fgr:EC_MAPI_CSPI}. When considering a water environment, the Gibbs free energy change of the decomposition reaction mentioned above decreased by 0.39~eV and becomes strongly negative. This decrease is due to hydration of \ch{MA+}/\ch{Cs+} and \ch{I-} ions.
The amount of enthalpy change during the designed two-step thermodynamic cycle for \ch{MAPbI3} dissolution in water corresponding to Eq.~(\ref{Eq:HC_MAPbI3_step0}) is negative (see $\Delta H$ in Fig.~\ref{fgr:EC_MAPI_CSPI}). This $\Delta H$ indicates the released energy during hydration process of \ch{MA+} and \ch{I-} in Eq.~(\ref{Eq:HC_MAPbI3_step2}) overcomes the energy consumed in Eq.~(\ref{Eq:HC_MAPbI3_step1}). It renders the dissolution \ch{MAPbI3} in water is an exothermic reaction. The entropy gain during the dissolution of \ch{MAPbI3} in water further lowers $\Delta G$ in Fig.~\ref{fgr:EC_MAPI_CSPI}, suggesting the dissolution process is thermodynamically preferable. Dissimilarly, a postive $\Delta H=0.28$~eV for \ch{CsPbI3} dissolving in water indicates the released energy during hydration process of \ch{Cs+} and \ch{I-} in Eq.~(\ref{Eq:HC_CsPbI3_step2}) is less than the energy consumed during Eq.~(\ref{Eq:HC_CsPbI3_step1}). This positive $\Delta H$ renders dissolution \ch{CsPbI3} in water belongs to an endothermic reaction. Interestingly, a large entropy enhancement of dissolution \ch{CsPbI3} in water brings down $\Delta G$ to a negative value, see in Fig.~\ref{fgr:EC_MAPI_CSPI}. This large entropy enhancement comes from the difference of a low entropy of orthrhombic \ch{CsPbI3} and a high entropy of \ch{Cs+}(aq). The relative high entropy of \ch{MAPbI3} and low entropy of \ch{CsPbI3} in Table.~\ref{Table:step1_MAPbI3} are consistent with the strong anharmonicity of \ch{MAPbI3} and weak anharmonicity of \ch{CsPbI3} recently disscussed in Ref.~\citenum{Zhu_EES_12_2019, Marronnier_ACSNN12_2018}. Compared with \ch{CsPbI3}, the large hydration enthalpy of \ch{MA+} ion and the entropy gain together direct the negative $\Delta G_\text{diss}$. In all, we can conclude that the large hydration enthalpies of \ch{MA+}/\ch{Cs+} and \ch{I-} and entropy enhancement as well as the low lattice energies of halide perovskite structures are responsible for the degradation of \ch{MAPbI3}/\ch{CsPbI3} in water.

So far, we did not discuss an ionic activity $a_\pm$ of CH$_3$NH$_3$I solution and its contribution $k_B T \ln (a_\pm)$ to a chemical potential term of an electrolyte when estimating $\tilde{\mu}^\circ_{T}$(aq). The actual chemical potential of an electrolyte is expressed as
\begin{equation}\label{Eq:Chem-potential-activity-aq}
	\mu_{T} \approx \tilde{\mu}^\circ_{T}(\text{aq}) + k_B T \ln (a_\pm).
\end{equation}
The mean ionic activity $a_\pm$ of CH$_3$NH$_3$I solution
\begin{equation}\label{Eq:Activity}
	a_\pm = (\gamma_\pm c/c^\circ)^2
\end{equation}
is determined by its molar concentration $c$ relative to the concentration in standard state $c^\circ = 1$~M and the mean ionic activity coefficient $\gamma_\pm$, which accounts for non-ideality of the solution. Thermodynamic data listed in Table~\ref{Table:step1_MAPbI3} correspond to $c = 1$~M. \citet{Bonner_JCSFT_77_1981} and \citet{ Belveze_IECR_43_2004} reported $\gamma_\pm$ of \ch{MACl} is 0.58 when $c = 1$~M. This value is taken as an estimation of $\gamma_\pm$ for \ch{MAI}. The additional term $k_\text{B} T \ln (a_\pm)$ for \ch{MAI} solution is estimated as $-0.028$~eV at $c=c^\circ$, which is relatively small considering $\Delta G_\text{diss}=-0.36$~eV. In a dilute solution limit ($c\ll 1$~M), the term $k_B T \ln (a_\pm)$ becomes negative and drives the water decomposition reaction of \ch{MAPbI3}\cite{Tenuta_SR_6_2016}. When decomposition proceeds to reach a finite concentration (\textit{e.g.}, $c=c^\circ$), the $\Delta G_\text{diss}$ is no longer governed by $k_B T \ln (a_\pm)$ but dominant by enthalpies and entropies of aqueous ions (see $\Delta H$ and $-T\Delta S$ of dissolution process in Fig.~\ref{fgr:EC_MAPI_CSPI}). Upon further dissolution, chemical potentials of the aqueous solution increase until $\Delta G_\text{diss} = 0$~eV, the solution is then saturated and the dissolution process  ceases.

\section{Conclusion}
The water instability of \ch{MAPbI3} is a major problem for commercialization of perovskite photovoltaics. Explanations of the underline mechanism are under debate. Here, we use \textit{ab initio} metadynamic method to reconstruct the free energy surface of the dissolution process of \ch{MAPbI3} in explicit water. The predictive power of metadynamics unravels the pathway of water dissolving \ch{MAPbI3} surface. One intermediate state and two transition states are identified during the initial dissolution process. The first transition state involves breaking of \ch{I-Pb} bond and formation of an intermediate state (\ch{I-MA} interactions) at the surface with a low energy barrier of 0.16~eV.  The second transition state corresponds to dissociation \ch{I-MA} interaction and hydration of the \ch{I-} ion with the energy barrier of 0.22~eV. In addition, using DFT calculations augmented with experimental data, the analysis of thermodynamics of \ch{MAPbI3} decomposition in water at a finite concentration indicates a negative Gibbs free energy change which suggests the spontaneity of water dissolution of \ch{MAPbI3}. It is worthy to mention that a large hydration enthalpy of \ch{MA+} and a entropy gain under aqueous condition direct the negative $\Delta G_\text{diss}$. Combined with the low energy barrier for ease of ions escaping from the \ch{MAPbI3} surface and spontaneous nature of dissolving in water, it can be explained why the water immediately destroys pristine \ch{MAPbI3}. We also analyze how \ch{CsPbI3} react with water from the thermodynamic point of view. It is found the Gibbs free energy change of dissolution \ch{CsPbI3} in water is also a negative value similar as the value of \ch{MAPbI3}, which is consistent with an experimental observation of \ch{CsPbI3} degradation in a moist environment. Compared with \ch{MAPbI3}, a large entropy enhancement of \ch{Cs+} dominates the negative $\Delta G_\text{diss}$ for \ch{CsPbI3} dissolution in water. Our explanation provides a deeper insight of the water instability of \ch{MAPbI3} and presents a perspective on improving the instability of perovskite photovoltaics in future.

\begin{suppinfo}

The following files are available free of charge.
\begin{itemize}
  \item Initial FES of \ch{MA+}: The initial basin of free energy surface for \ch{MA+} dissolution in water is plotted and compared with the initial basin of \ch{I-} dissolution in water.
\end{itemize}

\end{suppinfo}
\begin{acknowledgement}

Authors are indebted to Prof.~Claudine Katan, Dr.~Mika\"{e}l Kepenekian and Prof.~Xavier Rocquefelte from Universit\'{e} de Rennes~1 for advises on setting up the initial perovskite-water interface. The authors also thank Natural Sciences and Engineering Research Council of Canada under the Discovery Grant Programs RGPIN-2015-04518. All the calculations were performed using a Compute Canada infrastructure supported by the Canada Foundation for Innovation under the John R. Evans Leaders Fund program.

\end{acknowledgement}


\providecommand{\latin}[1]{#1}
\makeatletter
\providecommand{\doi}
  {\begingroup\let\do\@makeother\dospecials
  \catcode`\{=1 \catcode`\}=2 \doi@aux}
\providecommand{\doi@aux}[1]{\endgroup\texttt{#1}}
\makeatother
\providecommand*\mcitethebibliography{\thebibliography}
\csname @ifundefined\endcsname{endmcitethebibliography}
  {\let\endmcitethebibliography\endthebibliography}{}
\begin{mcitethebibliography}{73}
\providecommand*\natexlab[1]{#1}
\providecommand*\mciteSetBstSublistMode[1]{}
\providecommand*\mciteSetBstMaxWidthForm[2]{}
\providecommand*\mciteBstWouldAddEndPuncttrue
  {\def\EndOfBibitem{\unskip.}}
\providecommand*\mciteBstWouldAddEndPunctfalse
  {\let\EndOfBibitem\relax}
\providecommand*\mciteSetBstMidEndSepPunct[3]{}
\providecommand*\mciteSetBstSublistLabelBeginEnd[3]{}
\providecommand*\EndOfBibitem{}
\mciteSetBstSublistMode{f}
\mciteSetBstMaxWidthForm{subitem}{(\alph{mcitesubitemcount})}
\mciteSetBstSublistLabelBeginEnd
  {\mcitemaxwidthsubitemform\space}
  {\relax}
  {\relax}

\bibitem[Kojima \latin{et~al.}(2009)Kojima, Teshima, Shirai, and
  Miyasaka]{Kojima_JACS_131_2009}
Kojima,~A.; Teshima,~K.; Shirai,~Y.; Miyasaka,~T. Organometal Halide
  Perovskites as Visible-Light Sensitizers for Photovoltaic Cells. \emph{J. Am.
  Chem. Soc.} \textbf{2009}, \emph{131}, 6050--6051\relax
\mciteBstWouldAddEndPuncttrue
\mciteSetBstMidEndSepPunct{\mcitedefaultmidpunct}
{\mcitedefaultendpunct}{\mcitedefaultseppunct}\relax
\EndOfBibitem
\bibitem[Hailegnaw \latin{et~al.}(2015)Hailegnaw, Kirmayer, Edri, Hodes, and
  Cahen]{Hailegnaw_JPCL_6_2015}
Hailegnaw,~B.; Kirmayer,~S.; Edri,~E.; Hodes,~G.; Cahen,~D. Rain on
  Methylammonium Lead Iodide Based Perovskites: Possible Environmental Effects
  of Perovskite Solar Cells. \emph{J. Phys. Chem. Lett.} \textbf{2015},
  \emph{6}, 1543--1547\relax
\mciteBstWouldAddEndPuncttrue
\mciteSetBstMidEndSepPunct{\mcitedefaultmidpunct}
{\mcitedefaultendpunct}{\mcitedefaultseppunct}\relax
\EndOfBibitem
\bibitem[Li \latin{et~al.}(2015)Li, Xu, Wang, Wang, Xie, Yang, and
  Gao]{Li_JPCC_119_2015}
Li,~Y.; Xu,~X.; Wang,~C.; Wang,~C.; Xie,~F.; Yang,~J.; Gao,~Y. Degradation by
  Exposure of Coevaporated \ch{CH3NH3PbI3} Thin Films. \emph{J. Phys. Chem. C}
  \textbf{2015}, \emph{119}, 23996--24002\relax
\mciteBstWouldAddEndPuncttrue
\mciteSetBstMidEndSepPunct{\mcitedefaultmidpunct}
{\mcitedefaultendpunct}{\mcitedefaultseppunct}\relax
\EndOfBibitem
\bibitem[Ke \latin{et~al.}(2017)Ke, Walton, Lewis, Tedstone, O'Brien, Thomas,
  and Flavell]{Ke_CC_53_2017}
Ke,~J. C.-R.; Walton,~A.~S.; Lewis,~D.~J.; Tedstone,~A.; O'Brien,~P.;
  Thomas,~A.~G.; Flavell,~W.~R. \textit{In situ} Investigation of Degradation
  at Organometal Halide Perovskite Surfaces by X-ray Photoelectron Spectroscopy
  at Realistic Water Vapour Pressure. \emph{Chem. Commun.} \textbf{2017},
  \emph{53}, 5231--5234\relax
\mciteBstWouldAddEndPuncttrue
\mciteSetBstMidEndSepPunct{\mcitedefaultmidpunct}
{\mcitedefaultendpunct}{\mcitedefaultseppunct}\relax
\EndOfBibitem
\bibitem[Kakekhani \latin{et~al.}(2019)Kakekhani, Katti, and
  Rappe]{Kakekhani_APLM_7_2019}
Kakekhani,~A.; Katti,~R.~N.; Rappe,~A.~M. Water in Hybrid Perovskites: Bulk
  \ch{MAPbI3} Degradation \textit{via} Super-Hydrous State. \emph{APL Mater.}
  \textbf{2019}, \emph{7}, 041112\relax
\mciteBstWouldAddEndPuncttrue
\mciteSetBstMidEndSepPunct{\mcitedefaultmidpunct}
{\mcitedefaultendpunct}{\mcitedefaultseppunct}\relax
\EndOfBibitem
\bibitem[Philippe \latin{et~al.}(2015)Philippe, Park, Lindblad, Oscarsson,
  Ahmadi, Johansson, and Rensmo]{Philippe_CM_27_2015}
Philippe,~B.; Park,~B.-W.; Lindblad,~R.; Oscarsson,~J.; Ahmadi,~S.;
  Johansson,~E.~M.; Rensmo,~H. Chemical and Electronic Structure
  Characterization of Lead Halide Perovskites and Stability Behavior under
  Different Exposures A Photoelectron Spectroscopy Investigation. \emph{Chem.
  Mater.} \textbf{2015}, \emph{27}, 1720--1731\relax
\mciteBstWouldAddEndPuncttrue
\mciteSetBstMidEndSepPunct{\mcitedefaultmidpunct}
{\mcitedefaultendpunct}{\mcitedefaultseppunct}\relax
\EndOfBibitem
\bibitem[Niu \latin{et~al.}(2014)Niu, Li, Meng, Wang, Dong, and
  Qiu]{Niu_JMCA_2_2014}
Niu,~G.; Li,~W.; Meng,~F.; Wang,~L.; Dong,~H.; Qiu,~Y. Study on the Stability
  of \ch{CH3NH3PbI3} Films and the Effect of Post-Modification by Aluminum
  Oxide in All-Solid-State Hybrid Solar Cells. \emph{J. Mater. Chem. A}
  \textbf{2014}, \emph{2}, 705--710\relax
\mciteBstWouldAddEndPuncttrue
\mciteSetBstMidEndSepPunct{\mcitedefaultmidpunct}
{\mcitedefaultendpunct}{\mcitedefaultseppunct}\relax
\EndOfBibitem
\bibitem[Frost \latin{et~al.}(2014)Frost, Butler, Brivio, Hendon,
  Van~Schilfgaarde, and Walsh]{Frost_NL_14_2014}
Frost,~J.~M.; Butler,~K.~T.; Brivio,~F.; Hendon,~C.~H.; Van~Schilfgaarde,~M.;
  Walsh,~A. Atomistic Origins of High-Performance in Hybrid Halide Perovskite
  Solar Cells. \emph{Nano Lett.} \textbf{2014}, \emph{14}, 2584--2590\relax
\mciteBstWouldAddEndPuncttrue
\mciteSetBstMidEndSepPunct{\mcitedefaultmidpunct}
{\mcitedefaultendpunct}{\mcitedefaultseppunct}\relax
\EndOfBibitem
\bibitem[Wei and Hu(2016)Wei, and Hu]{Wei_IJER_41_2016}
Wei,~W.; Hu,~Y.~H. Catalytic Role of \ch{H2O} in Degradation of
  Inorganic-Organic Perovskite (\ch{CH3NH3PbI3}) in Air. \emph{Int. J. Energy
  Res.} \textbf{2016}, \emph{41}, 1063--1069\relax
\mciteBstWouldAddEndPuncttrue
\mciteSetBstMidEndSepPunct{\mcitedefaultmidpunct}
{\mcitedefaultendpunct}{\mcitedefaultseppunct}\relax
\EndOfBibitem
\bibitem[McLeod and Liu(2018)McLeod, and Liu]{McLeod_JPCL_9_2018}
McLeod,~J.~A.; Liu,~L. Prospects for Mitigating Intrinsic Organic Decomposition
  in Methylammonium Lead Triiodide Perovskite. \emph{J. Phys. Chem. Lett.}
  \textbf{2018}, \emph{9}, 2411--2417\relax
\mciteBstWouldAddEndPuncttrue
\mciteSetBstMidEndSepPunct{\mcitedefaultmidpunct}
{\mcitedefaultendpunct}{\mcitedefaultseppunct}\relax
\EndOfBibitem
\bibitem[Yang \latin{et~al.}(2018)Yang, Yuan, Liu, Braun, Li, Tang, Gao, Duan,
  Fahlman, and Bao]{Yang_AMI_10_2018}
Yang,~J.; Yuan,~Z.; Liu,~X.; Braun,~S.; Li,~Y.; Tang,~J.; Gao,~F.; Duan,~C.;
  Fahlman,~M.; Bao,~Q. Oxygen- and Water-Induced Energetics Degradation in
  Organometal Halide Perovskites. \emph{ACS Appl. Mater. Interfaces}
  \textbf{2018}, \emph{10}, 16225--16230\relax
\mciteBstWouldAddEndPuncttrue
\mciteSetBstMidEndSepPunct{\mcitedefaultmidpunct}
{\mcitedefaultendpunct}{\mcitedefaultseppunct}\relax
\EndOfBibitem
\bibitem[Kosasih and Ducati(2018)Kosasih, and Ducati]{Kosasih_NE_47_2018}
Kosasih,~F.~U.; Ducati,~C. Characterising Degradation of Perovskite Solar Cells
  Through In-Situ and Operando Electron Microscopy. \emph{Nano Energy}
  \textbf{2018}, \emph{47}, 243--256\relax
\mciteBstWouldAddEndPuncttrue
\mciteSetBstMidEndSepPunct{\mcitedefaultmidpunct}
{\mcitedefaultendpunct}{\mcitedefaultseppunct}\relax
\EndOfBibitem
\bibitem[Aristidou \latin{et~al.}(2017)Aristidou, Eames, Islam, and
  Haque]{Aristidou_JMCA_5_2017}
Aristidou,~N.; Eames,~C.; Islam,~M.~S.; Haque,~S.~A. Insights into the
  Increased Degradation Rate of \ch{CH3NH3PbI3} Solar Cells in Combined Water
  and \ch{O2} Environments. \emph{J. Mater. Chem. A} \textbf{2017}, \emph{5},
  25469--25475\relax
\mciteBstWouldAddEndPuncttrue
\mciteSetBstMidEndSepPunct{\mcitedefaultmidpunct}
{\mcitedefaultendpunct}{\mcitedefaultseppunct}\relax
\EndOfBibitem
\bibitem[Smecca \latin{et~al.}(2016)Smecca, Numata, Deretzis, Pellegrino,
  Boninelli, Miyasaka, Magna, and Alberti]{Smecca_PCCP_18_2016}
Smecca,~E.; Numata,~Y.; Deretzis,~I.; Pellegrino,~G.; Boninelli,~S.;
  Miyasaka,~T.; Magna,~A.~L.; Alberti,~A. Stability of Solution-Processed
  \ch{MAPbI3} and \ch{FAPbI3} Layers. \emph{Phys. Chem. Chem. Phys.}
  \textbf{2016}, \emph{18}, 13413--13422\relax
\mciteBstWouldAddEndPuncttrue
\mciteSetBstMidEndSepPunct{\mcitedefaultmidpunct}
{\mcitedefaultendpunct}{\mcitedefaultseppunct}\relax
\EndOfBibitem
\bibitem[Saidaminov \latin{et~al.}(2018)Saidaminov, Kim, Jain,
  Quintero-Bermudez, Tan, Long, Tan, Johnston, Zhao, Voznyy, and
  Sargent]{Saidaminov_NE_3_2018}
Saidaminov,~M.~I.; Kim,~J.; Jain,~A.; Quintero-Bermudez,~R.; Tan,~H.; Long,~G.;
  Tan,~F.; Johnston,~A.; Zhao,~Y.; Voznyy,~O. \latin{et~al.}  Suppression of
  Atomic Vacancies via Incorporation of Isovalent Small Ions to Increase the
  Stability of Halide Perovskite Solar Cells in Ambient Air. \emph{Nat. Energy}
  \textbf{2018}, \emph{3}, 648--654\relax
\mciteBstWouldAddEndPuncttrue
\mciteSetBstMidEndSepPunct{\mcitedefaultmidpunct}
{\mcitedefaultendpunct}{\mcitedefaultseppunct}\relax
\EndOfBibitem
\bibitem[Lin \latin{et~al.}(2015)Lin, Lyu, Hunston, Kim, Wan, Stanley, and
  Gu]{Lin_ISOP_2015}
Lin,~C.-C.; Lyu,~Y.; Hunston,~D.~L.; Kim,~J.~H.; Wan,~K.-T.; Stanley,~D.~L.;
  Gu,~X. Cracking and Delamination Behaviors of Photovoltaic Backsheet after
  Accelerated Laboratory Weathering. Reliability of Photovoltaic Cells,
  Modules, Components, and Systems VIII. 2015; p 956304\relax
\mciteBstWouldAddEndPuncttrue
\mciteSetBstMidEndSepPunct{\mcitedefaultmidpunct}
{\mcitedefaultendpunct}{\mcitedefaultseppunct}\relax
\EndOfBibitem
\bibitem[Mosconi \latin{et~al.}(2015)Mosconi, Azpiroz, and
  De~Angelis]{Mosconi_CM_27_2015}
Mosconi,~E.; Azpiroz,~J.~M.; De~Angelis,~F. \textit{Ab Initio} Molecular
  Dynamics Simulations of Methylammonium Lead Iodide Perovskite Degradation by
  Water. \emph{Chem. Mater.} \textbf{2015}, \emph{27}, 4885--4892\relax
\mciteBstWouldAddEndPuncttrue
\mciteSetBstMidEndSepPunct{\mcitedefaultmidpunct}
{\mcitedefaultendpunct}{\mcitedefaultseppunct}\relax
\EndOfBibitem
\bibitem[Caddeo \latin{et~al.}(2018)Caddeo, Marongiu, Meloni, Filippetti,
  Quochi, Saba, and Mattoni]{Caddeo_AMI_2018}
Caddeo,~C.; Marongiu,~D.; Meloni,~S.; Filippetti,~A.; Quochi,~F.; Saba,~M.;
  Mattoni,~A. Hydrophilicity and Water Contact Angle on Methylammonium Lead
  Iodide. \emph{Adv. Mater. Interfaces} \textbf{2018}, 1801173\relax
\mciteBstWouldAddEndPuncttrue
\mciteSetBstMidEndSepPunct{\mcitedefaultmidpunct}
{\mcitedefaultendpunct}{\mcitedefaultseppunct}\relax
\EndOfBibitem
\bibitem[Caddeo \latin{et~al.}(2017)Caddeo, Saba, Meloni, Filippetti, and
  Mattoni]{Caddeo_AN_11_2017}
Caddeo,~C.; Saba,~M.~I.; Meloni,~S.; Filippetti,~A.; Mattoni,~A. Collective
  Molecular Mechanisms in the \ch{CH3NH3PbI3} Dissolution by Liquid Water.
  \emph{ACS nano} \textbf{2017}, \emph{11}, 9183--9190\relax
\mciteBstWouldAddEndPuncttrue
\mciteSetBstMidEndSepPunct{\mcitedefaultmidpunct}
{\mcitedefaultendpunct}{\mcitedefaultseppunct}\relax
\EndOfBibitem
\bibitem[Park \latin{et~al.}(2017)Park, Chang, Lee, Park, Ahn, and
  Nam]{Park_NE_2_2017}
Park,~S.; Chang,~W.~J.; Lee,~C.~W.; Park,~S.; Ahn,~H.-Y.; Nam,~K.~T.
  Photocatalytic Hydrogen Generation from Hydriodic Acid Using Methylammonium
  Lead Iodide in Dynamic Equilibrium with Aqueous Solution. \emph{Nat. Energy}
  \textbf{2017}, \emph{2}, 16185\relax
\mciteBstWouldAddEndPuncttrue
\mciteSetBstMidEndSepPunct{\mcitedefaultmidpunct}
{\mcitedefaultendpunct}{\mcitedefaultseppunct}\relax
\EndOfBibitem
\bibitem[Liu \latin{et~al.}(2011)Liu, Laio, and Michaelides]{Liu_PCCP_13_2011}
Liu,~L.-M.; Laio,~A.; Michaelides,~A. Initial Stages of Salt Crystal
  Dissolution Determined with \textit{Ab Initio} Molecular Dynamics.
  \emph{Phys. Chem. Chem. Phys.} \textbf{2011}, \emph{13}, 13162--13166\relax
\mciteBstWouldAddEndPuncttrue
\mciteSetBstMidEndSepPunct{\mcitedefaultmidpunct}
{\mcitedefaultendpunct}{\mcitedefaultseppunct}\relax
\EndOfBibitem
\bibitem[Chen \latin{et~al.}(2014)Chen, Reischl, Spijker, Holmberg, Laasonen,
  and Foster]{Chen_PCCP_16_2014}
Chen,~J.-C.; Reischl,~B.; Spijker,~P.; Holmberg,~N.; Laasonen,~K.;
  Foster,~A.~S. \textit{Ab Initio} Kinetic Monte Carlo Simulations of
  Dissolution at the NaCl-Water Interface. \emph{Phys. Chem. Chem. Phys.}
  \textbf{2014}, \emph{16}, 22545--22554\relax
\mciteBstWouldAddEndPuncttrue
\mciteSetBstMidEndSepPunct{\mcitedefaultmidpunct}
{\mcitedefaultendpunct}{\mcitedefaultseppunct}\relax
\EndOfBibitem
\bibitem[Laio and Parrinello(2002)Laio, and Parrinello]{Laio_PNAS_99_2002}
Laio,~A.; Parrinello,~M. Escaping Free-Energy Minima. \emph{Proc. Natl. Acad.
  Sci. U.S.A.} \textbf{2002}, \emph{99}, 12562--12566\relax
\mciteBstWouldAddEndPuncttrue
\mciteSetBstMidEndSepPunct{\mcitedefaultmidpunct}
{\mcitedefaultendpunct}{\mcitedefaultseppunct}\relax
\EndOfBibitem
\bibitem[Iannuzzi \latin{et~al.}(2003)Iannuzzi, Laio, and
  Parrinello]{Lannuzzi_PRL_90_2003}
Iannuzzi,~M.; Laio,~A.; Parrinello,~M. Efficient Exploration of Reactive
  Potential Energy Surfaces Using Car-Parrinello Molecular Dynamics.
  \emph{Phys. Rev. Lett.} \textbf{2003}, \emph{90}, 238302\relax
\mciteBstWouldAddEndPuncttrue
\mciteSetBstMidEndSepPunct{\mcitedefaultmidpunct}
{\mcitedefaultendpunct}{\mcitedefaultseppunct}\relax
\EndOfBibitem
\bibitem[Bucko(2008)]{Bucko_JPCM_20_2008}
Bucko,~T. \textit{Ab Initio} Calculations of Free-Energy Reaction Barriers.
  \emph{J. Phys.: Condens. Matter} \textbf{2008}, \emph{20}, 064211\relax
\mciteBstWouldAddEndPuncttrue
\mciteSetBstMidEndSepPunct{\mcitedefaultmidpunct}
{\mcitedefaultendpunct}{\mcitedefaultseppunct}\relax
\EndOfBibitem
\bibitem[Kohn and Sham(1965)Kohn, and Sham]{Kohn_PR_140_1965}
Kohn,~W.; Sham,~L.~J. Self-Consistent Equations Including Exchange and
  Correlation Effects. \emph{Phys. Rev.} \textbf{1965}, \emph{140}, A1133\relax
\mciteBstWouldAddEndPuncttrue
\mciteSetBstMidEndSepPunct{\mcitedefaultmidpunct}
{\mcitedefaultendpunct}{\mcitedefaultseppunct}\relax
\EndOfBibitem
\bibitem[Perdew \latin{et~al.}(1996)Perdew, Burke, and
  Ernzerhof]{Perdew_PRL_77_1996}
Perdew,~J.~P.; Burke,~K.; Ernzerhof,~M. Generalized Gradient Approximation Made
  Simple. \emph{Phys. Rev. Lett.} \textbf{1996}, \emph{77}, 3865\relax
\mciteBstWouldAddEndPuncttrue
\mciteSetBstMidEndSepPunct{\mcitedefaultmidpunct}
{\mcitedefaultendpunct}{\mcitedefaultseppunct}\relax
\EndOfBibitem
\bibitem[Wang \latin{et~al.}(2013)Wang, Gould, Dobson, Zhang, Yang, Yao, and
  Zhao]{Wang_PCCP_16_2013}
Wang,~Y.; Gould,~T.; Dobson,~J.~F.; Zhang,~H.; Yang,~H.; Yao,~X.; Zhao,~H.
  Density Functional Theory Analysis of Structural and Electronic Properties of
  Orthorhombic Perovskite \ch{CH3NH3PbI3}. \emph{Phys. Chem. Chem. Phys.}
  \textbf{2013}, \emph{16}, 1424--1429\relax
\mciteBstWouldAddEndPuncttrue
\mciteSetBstMidEndSepPunct{\mcitedefaultmidpunct}
{\mcitedefaultendpunct}{\mcitedefaultseppunct}\relax
\EndOfBibitem
\bibitem[Zheng and Rubel(2018)Zheng, and Rubel]{Zheng_9_JPCL_2018}
Zheng,~C.; Rubel,~O. Aziridinium Lead Iodide: A Stable, Low-Band-Gap Hybrid
  Halide Perovskite for Photovoltaics. \emph{J. Phys. Chem. Lett.}
  \textbf{2018}, \emph{9}, 874--880\relax
\mciteBstWouldAddEndPuncttrue
\mciteSetBstMidEndSepPunct{\mcitedefaultmidpunct}
{\mcitedefaultendpunct}{\mcitedefaultseppunct}\relax
\EndOfBibitem
\bibitem[Li and Yang(2019)Li, and Yang]{Li_EES_12_2019}
Li,~Y.; Yang,~K. High-Throughput Computational Design of Organic-Inorganic
  Hybrid Halide Semiconductors beyond Perovskites for Optoelectronics.
  \emph{Energy Environ. Sci.} \textbf{2019}, \emph{12}, 2233--2243\relax
\mciteBstWouldAddEndPuncttrue
\mciteSetBstMidEndSepPunct{\mcitedefaultmidpunct}
{\mcitedefaultendpunct}{\mcitedefaultseppunct}\relax
\EndOfBibitem
\bibitem[Klime{\v{s}} \latin{et~al.}(2009)Klime{\v{s}}, Bowler, and
  Michaelides]{Klimevs_JPCM_22_2009}
Klime{\v{s}},~J.; Bowler,~D.~R.; Michaelides,~A. Chemical Accuracy for the Van
  Der Waals Density Functional. \emph{J Phys. Condens. Matter} \textbf{2009},
  \emph{22}, 022201\relax
\mciteBstWouldAddEndPuncttrue
\mciteSetBstMidEndSepPunct{\mcitedefaultmidpunct}
{\mcitedefaultendpunct}{\mcitedefaultseppunct}\relax
\EndOfBibitem
\bibitem[Grimme \latin{et~al.}(2010)Grimme, Antony, Ehrlich, and
  Krieg]{Grimme_JCP_132_2010}
Grimme,~S.; Antony,~J.; Ehrlich,~S.; Krieg,~H. A Consistent and Accurate
  \textit{Ab Initio} Parametrization of Density Functional Dispersion
  Correction (DFT-D) for the 94 Elements H-Pu. \emph{J. Chem. Phys.}
  \textbf{2010}, \emph{132}, 154104\relax
\mciteBstWouldAddEndPuncttrue
\mciteSetBstMidEndSepPunct{\mcitedefaultmidpunct}
{\mcitedefaultendpunct}{\mcitedefaultseppunct}\relax
\EndOfBibitem
\bibitem[Kresse and Furthm{\"u}ller(1996)Kresse, and
  Furthm{\"u}ller]{Kresse_PRB_54_1996}
Kresse,~G.; Furthm{\"u}ller,~J. Efficient Iterative Schemes for \textit{Ab
  Initio} Total-Energy Calculations Using A Plane-Wave Basis Set. \emph{Phys.
  Rev. B} \textbf{1996}, \emph{54}, 11169\relax
\mciteBstWouldAddEndPuncttrue
\mciteSetBstMidEndSepPunct{\mcitedefaultmidpunct}
{\mcitedefaultendpunct}{\mcitedefaultseppunct}\relax
\EndOfBibitem
\bibitem[Kresse and Joubert(1999)Kresse, and Joubert]{Kresse_PRB_59_1999}
Kresse,~G.; Joubert,~D. From Ultrasoft Pseudopotentials to the Projector
  Augmented-Wave Method. \emph{Phys. Rev. B} \textbf{1999}, \emph{59},
  1758\relax
\mciteBstWouldAddEndPuncttrue
\mciteSetBstMidEndSepPunct{\mcitedefaultmidpunct}
{\mcitedefaultendpunct}{\mcitedefaultseppunct}\relax
\EndOfBibitem
\bibitem[Bl{\"o}chl(1994)]{Blochl_PRB_50_1994}
Bl{\"o}chl,~P. Projector Augmented-Wave Method. \emph{Phys. Rev. B}
  \textbf{1994}, \emph{50}, 17953\relax
\mciteBstWouldAddEndPuncttrue
\mciteSetBstMidEndSepPunct{\mcitedefaultmidpunct}
{\mcitedefaultendpunct}{\mcitedefaultseppunct}\relax
\EndOfBibitem
\bibitem[Zheng and Rubel(2017)Zheng, and Rubel]{Zheng_JPCC_131_2017}
Zheng,~C.; Rubel,~O. Ionization Energy as a Stability Criterion for Halide
  Perovskites. \emph{J. Phys. Chem. C} \textbf{2017}, \emph{121},
  11977--11984\relax
\mciteBstWouldAddEndPuncttrue
\mciteSetBstMidEndSepPunct{\mcitedefaultmidpunct}
{\mcitedefaultendpunct}{\mcitedefaultseppunct}\relax
\EndOfBibitem
\bibitem[Monkhorst and Pack(1976)Monkhorst, and Pack]{Monkhorst_PRB_13_1976}
Monkhorst,~H.~J.; Pack,~J.~D. Special Points for Brillouin-Zone Integrations.
  \emph{Phys. Rev. B} \textbf{1976}, \emph{13}, 5188--5192\relax
\mciteBstWouldAddEndPuncttrue
\mciteSetBstMidEndSepPunct{\mcitedefaultmidpunct}
{\mcitedefaultendpunct}{\mcitedefaultseppunct}\relax
\EndOfBibitem
\bibitem[Zheng \latin{et~al.}(2018)Zheng, Yu, and Rubel]{Zheng_PRM_2_2018}
Zheng,~C.; Yu,~S.; Rubel,~O. Structural Dynamics in Hybrid Halide Perovskites:
  Bulk Rashba Splitting, Spin Texture, and Carrier Localization. \emph{Phys.
  Rev. Materials} \textbf{2018}, \emph{2}, 114604\relax
\mciteBstWouldAddEndPuncttrue
\mciteSetBstMidEndSepPunct{\mcitedefaultmidpunct}
{\mcitedefaultendpunct}{\mcitedefaultseppunct}\relax
\EndOfBibitem
\bibitem[Whitfield \latin{et~al.}(2016)Whitfield, Herron, Guise, Page, Cheng,
  Milas, and Crawford]{Whitfield_SR_6_2016}
Whitfield,~P.; Herron,~N.; Guise,~W.; Page,~K.; Cheng,~Y.; Milas,~I.;
  Crawford,~M. Structures, Phase Transitions and Tricritical Behavior of the
  Hybrid Perovskite Methyl Ammonium Lead Iodide. \emph{Sci. Rep.}
  \textbf{2016}, \emph{6}, 35685\relax
\mciteBstWouldAddEndPuncttrue
\mciteSetBstMidEndSepPunct{\mcitedefaultmidpunct}
{\mcitedefaultendpunct}{\mcitedefaultseppunct}\relax
\EndOfBibitem
\bibitem[She \latin{et~al.}(2015)She, Liu, and Zhong]{She_ACS_10_2015}
She,~L.; Liu,~M.; Zhong,~D. Atomic Structures of \ch{CH3NH3PbI3} (001)
  Surfaces. \emph{ACS Nano} \textbf{2015}, \emph{10}, 1126--1131\relax
\mciteBstWouldAddEndPuncttrue
\mciteSetBstMidEndSepPunct{\mcitedefaultmidpunct}
{\mcitedefaultendpunct}{\mcitedefaultseppunct}\relax
\EndOfBibitem
\bibitem[Ohmann \latin{et~al.}(2015)Ohmann, Ono, Kim, Lin, Lee, Li, Park, and
  Qi]{Ohmann_JACS_137_2015}
Ohmann,~R.; Ono,~L.~K.; Kim,~H.-S.; Lin,~H.; Lee,~M.~V.; Li,~Y.; Park,~N.-G.;
  Qi,~Y. Real-Space Imaging of the Atomic Structure of Organic--Inorganic
  Perovskite. \emph{J. Am. Chem. Soc.} \textbf{2015}, \emph{137},
  16049--16054\relax
\mciteBstWouldAddEndPuncttrue
\mciteSetBstMidEndSepPunct{\mcitedefaultmidpunct}
{\mcitedefaultendpunct}{\mcitedefaultseppunct}\relax
\EndOfBibitem
\bibitem[Nos{\'e}(1984)]{Nose_JCP_81_1984}
Nos{\'e},~S. A Unified Formulation of the Constant Temperature Molecular
  Dynamics Methods. \emph{J. Chem. Phys.} \textbf{1984}, \emph{81},
  511--519\relax
\mciteBstWouldAddEndPuncttrue
\mciteSetBstMidEndSepPunct{\mcitedefaultmidpunct}
{\mcitedefaultendpunct}{\mcitedefaultseppunct}\relax
\EndOfBibitem
\bibitem[Hoover(1985)]{Hoover_PRA_31_1985}
Hoover,~W.~G. Canonical Dynamics: Equilibrium Phase-Space Distributions.
  \emph{Phys. Rev. A} \textbf{1985}, \emph{31}, 1695\relax
\mciteBstWouldAddEndPuncttrue
\mciteSetBstMidEndSepPunct{\mcitedefaultmidpunct}
{\mcitedefaultendpunct}{\mcitedefaultseppunct}\relax
\EndOfBibitem
\bibitem[Ensing \latin{et~al.}(2005)Ensing, Laio, Parrinello, and
  Klein]{Ensing_PJCB_109_2005}
Ensing,~B.; Laio,~A.; Parrinello,~M.; Klein,~M.~L. A Recipe for the Computation
  of the Free Energy Barrier and the Lowest Free Energy Path of Concerted
  Reactions. \emph{J. Phys. Chem. B} \textbf{2005}, \emph{109},
  6676--6687\relax
\mciteBstWouldAddEndPuncttrue
\mciteSetBstMidEndSepPunct{\mcitedefaultmidpunct}
{\mcitedefaultendpunct}{\mcitedefaultseppunct}\relax
\EndOfBibitem
\bibitem[Karmakar and Chandra(2015)Karmakar, and
  Chandra]{Karmakar_JPCB_119_2015}
Karmakar,~A.; Chandra,~A. Water in Hydration Shell of An Iodide Ion: Structure
  and Dynamics of Solute-Water Hydrogen Bonds and Vibrational Spectral
  Diffusion from First-Principles Simulations. \emph{J. Phys. Chem. C}
  \textbf{2015}, \emph{119}, 8561--8572\relax
\mciteBstWouldAddEndPuncttrue
\mciteSetBstMidEndSepPunct{\mcitedefaultmidpunct}
{\mcitedefaultendpunct}{\mcitedefaultseppunct}\relax
\EndOfBibitem
\bibitem[Bonomi \latin{et~al.}(2009)Bonomi, Branduardi, Bussi, Camilloni,
  Provasi, Raiteri, Donadio, Marinelli, Pietrucci, and
  Broglia]{Bonomi_CPC_180_2009}
Bonomi,~M.; Branduardi,~D.; Bussi,~G.; Camilloni,~C.; Provasi,~D.; Raiteri,~P.;
  Donadio,~D.; Marinelli,~F.; Pietrucci,~F.; Broglia,~R.~A. PLUMED: A Portable
  Plugin for Free-Energy Calculations with Molecular Dynamics. \emph{Comput.
  Phys. Commun.} \textbf{2009}, \emph{180}, 1961--1972\relax
\mciteBstWouldAddEndPuncttrue
\mciteSetBstMidEndSepPunct{\mcitedefaultmidpunct}
{\mcitedefaultendpunct}{\mcitedefaultseppunct}\relax
\EndOfBibitem
\bibitem[Momma and Izumi(2011)Momma, and Izumi]{Momma_JAC_44_2011}
Momma,~K.; Izumi,~F. VESTA 3 for Three-Dimensional Visualization of Crystal,
  Volumetric and Morphology Data. \emph{J. Appl. Crystallogr.} \textbf{2011},
  \emph{44}, 1272--1276\relax
\mciteBstWouldAddEndPuncttrue
\mciteSetBstMidEndSepPunct{\mcitedefaultmidpunct}
{\mcitedefaultendpunct}{\mcitedefaultseppunct}\relax
\EndOfBibitem
\bibitem[Markovich \latin{et~al.}(1991)Markovich, Giniger, Levin, and
  Cheshnovsky]{Markovich_JCP_95_1991}
Markovich,~G.; Giniger,~R.; Levin,~M.; Cheshnovsky,~O. Photoelectron
  Spectroscopy of Iodine Anion Solvated in Water Clusters. \emph{J. Chem.
  Phys.} \textbf{1991}, \emph{95}, 9416--9419\relax
\mciteBstWouldAddEndPuncttrue
\mciteSetBstMidEndSepPunct{\mcitedefaultmidpunct}
{\mcitedefaultendpunct}{\mcitedefaultseppunct}\relax
\EndOfBibitem
\bibitem[Lin \latin{et~al.}(2018)Lin, Lai, Dou, Kley, Chen, Peng, Sun, Lu,
  Hawks, and Xie]{Lin_NM_17_2018}
Lin,~J.; Lai,~M.; Dou,~L.; Kley,~C.~S.; Chen,~H.; Peng,~F.; Sun,~J.; Lu,~D.;
  Hawks,~S.~A.; Xie,~C. Thermochromic Halide Perovskite Solar Cells. \emph{Nat.
  Mater.} \textbf{2018}, \emph{17}, 261--267\relax
\mciteBstWouldAddEndPuncttrue
\mciteSetBstMidEndSepPunct{\mcitedefaultmidpunct}
{\mcitedefaultendpunct}{\mcitedefaultseppunct}\relax
\EndOfBibitem
\bibitem[Ohtaki \latin{et~al.}(1988)Ohtaki, Fukushima, Hayakawa, and
  Okada]{Ohtaki_PAC_60_1988}
Ohtaki,~H.; Fukushima,~N.; Hayakawa,~E.; Okada,~I. Dissolution Process of
  Sodium Chloride Crystal in Water. \emph{Pure Appl. Chem.} \textbf{1988},
  \emph{60}, 1321--1324\relax
\mciteBstWouldAddEndPuncttrue
\mciteSetBstMidEndSepPunct{\mcitedefaultmidpunct}
{\mcitedefaultendpunct}{\mcitedefaultseppunct}\relax
\EndOfBibitem
\bibitem[Yang \latin{et~al.}(2005)Yang, Meng, Xu, Wang, and
  Gao]{Yang_PRE_72_2005}
Yang,~Y.; Meng,~S.; Xu,~L.~F.; Wang,~E.~G.; Gao,~S. Dissolution Dynamics of
  NaCl Nanocrystal in Liquid Water. \emph{Phys. Rev. E} \textbf{2005},
  \emph{72}, 012602\relax
\mciteBstWouldAddEndPuncttrue
\mciteSetBstMidEndSepPunct{\mcitedefaultmidpunct}
{\mcitedefaultendpunct}{\mcitedefaultseppunct}\relax
\EndOfBibitem
\bibitem[Fedotova and Kruchinin(2012)Fedotova, and
  Kruchinin]{Fedotova_RCB_61_2012}
Fedotova,~M.; Kruchinin,~S. Hydration of Methylamine and Methylammonium Ion:
  Structural and Thermodynamic Properties from the Data of the Integral
  Equation Method in the RISM Approximation. \emph{Russ. Chem. Bull.}
  \textbf{2012}, \emph{61}, 240--247\relax
\mciteBstWouldAddEndPuncttrue
\mciteSetBstMidEndSepPunct{\mcitedefaultmidpunct}
{\mcitedefaultendpunct}{\mcitedefaultseppunct}\relax
\EndOfBibitem
\bibitem[Delugas \latin{et~al.}(2015)Delugas, Filippetti, and
  Mattoni]{Delugas_PRB_92_2015}
Delugas,~P.; Filippetti,~A.; Mattoni,~A. Methylammonium Fragmentation in Amines
  as Source of Localized Trap Levels and the Healing Role of Cl in Hybrid
  Lead-Iodide Perovskites. \emph{Phys. Rev. B} \textbf{2015}, \emph{92},
  045301\relax
\mciteBstWouldAddEndPuncttrue
\mciteSetBstMidEndSepPunct{\mcitedefaultmidpunct}
{\mcitedefaultendpunct}{\mcitedefaultseppunct}\relax
\EndOfBibitem
\bibitem[Fan \latin{et~al.}(2017)Fan, Xiao, Wang, Zhao, Lin, Cheng, Lee, Wang,
  Feng, and Goddard~III]{Fan_JOULE_1_2017}
Fan,~Z.; Xiao,~H.; Wang,~Y.; Zhao,~Z.; Lin,~Z.; Cheng,~H.-C.; Lee,~S.-J.;
  Wang,~G.; Feng,~Z.; Goddard~III,~W.~A. Layer-by-Layer Degradation of
  Methylammonium Lead Tri-iodide Perovskite Microplates. \emph{Joule}
  \textbf{2017}, \emph{1}, 548--562\relax
\mciteBstWouldAddEndPuncttrue
\mciteSetBstMidEndSepPunct{\mcitedefaultmidpunct}
{\mcitedefaultendpunct}{\mcitedefaultseppunct}\relax
\EndOfBibitem
\bibitem[Sholl and Steckel(2011)Sholl, and Steckel]{Sholl_2011}
Sholl,~D.; Steckel,~J.~A. \emph{Density Functional Theory: a Practical
  Introduction}; John Wiley \& Sons: Hoboken, New Jersey, 2011\relax
\mciteBstWouldAddEndPuncttrue
\mciteSetBstMidEndSepPunct{\mcitedefaultmidpunct}
{\mcitedefaultendpunct}{\mcitedefaultseppunct}\relax
\EndOfBibitem
\bibitem[Tenuta \latin{et~al.}(2016)Tenuta, Zheng, and Rubel]{Tenuta_SR_6_2016}
Tenuta,~E.; Zheng,~C.; Rubel,~O. Thermodynamic Origin of Instability in Hybrid
  Halide Perovskites. \emph{Sci. Rep.} \textbf{2016}, \emph{6}, 37654\relax
\mciteBstWouldAddEndPuncttrue
\mciteSetBstMidEndSepPunct{\mcitedefaultmidpunct}
{\mcitedefaultendpunct}{\mcitedefaultseppunct}\relax
\EndOfBibitem
\bibitem[Kye \latin{et~al.}(2019)Kye, Yu, Jong, Ri, Kim, Choe, Hong, Li,
  Wilson, and Walsh]{Kye_JPCC_2019}
Kye,~Y.-H.; Yu,~C.-J.; Jong,~U.-G.; Ri,~K.-C.; Kim,~J.-S.; Choe,~S.-H.;
  Hong,~S.-N.; Li,~S.; Wilson,~J.~N.; Walsh,~A. Vacancy-Driven Stabilization of
  the Cubic Perovskite Polymorph of \ch{CsPbI3}. \emph{J. Phys. Chem. C}
  \textbf{2019}, \emph{123}, 9735--9744\relax
\mciteBstWouldAddEndPuncttrue
\mciteSetBstMidEndSepPunct{\mcitedefaultmidpunct}
{\mcitedefaultendpunct}{\mcitedefaultseppunct}\relax
\EndOfBibitem
\bibitem[Onoda-Yamamuro \latin{et~al.}(1990)Onoda-Yamamuro, Matsuo, and
  Suga]{Onoda_JPCS_51_1990}
Onoda-Yamamuro,~N.; Matsuo,~T.; Suga,~H. Calorimetric and IR Spectroscopic
  Studies of Phase Transitions in Methylammonium Trihalogenoplumbates (II).
  \emph{J. Phys. Chem. Solids} \textbf{1990}, \emph{51}, 1383--1395\relax
\mciteBstWouldAddEndPuncttrue
\mciteSetBstMidEndSepPunct{\mcitedefaultmidpunct}
{\mcitedefaultendpunct}{\mcitedefaultseppunct}\relax
\EndOfBibitem
\bibitem[Ong \latin{et~al.}(2015)Ong, Cholia, Jain, Brafman, Gunter, Ceder, and
  Persson]{Ong_CMS_97_2015}
Ong,~S.~P.; Cholia,~S.; Jain,~A.; Brafman,~M.; Gunter,~D.; Ceder,~G.;
  Persson,~K.~A. The Materials Application Programming Interface (API): a
  Simple, Flexible and Efficient API for Materials Data Based on
  REpresentational State Transfer (REST) Principles. \emph{Comput. Mater. Sci.}
  \textbf{2015}, \emph{97}, 209--215\relax
\mciteBstWouldAddEndPuncttrue
\mciteSetBstMidEndSepPunct{\mcitedefaultmidpunct}
{\mcitedefaultendpunct}{\mcitedefaultseppunct}\relax
\EndOfBibitem
\bibitem[Hinuma \latin{et~al.}(2017)Hinuma, Pizzi, Kumagai, Oba, and
  Tanaka]{Hinuma_CMS_128_2017}
Hinuma,~Y.; Pizzi,~G.; Kumagai,~Y.; Oba,~F.; Tanaka,~I. Band Structure Diagram
  Paths Based on Crystallography. \emph{Comput. Mater. Sci.} \textbf{2017},
  \emph{128}, 140--184\relax
\mciteBstWouldAddEndPuncttrue
\mciteSetBstMidEndSepPunct{\mcitedefaultmidpunct}
{\mcitedefaultendpunct}{\mcitedefaultseppunct}\relax
\EndOfBibitem
\bibitem[Yamamuro \latin{et~al.}(1986)Yamamuro, Oguni, Matsuo, and
  Suga]{Yamamuro_JCT_18_1986}
Yamamuro,~O.; Oguni,~M.; Matsuo,~T.; Suga,~H. Calorimetric and Dilatometric
  Studies on the Phase Transitions of Crystalline \ch{CH3NH3I}. \emph{J. Chem.
  Thermodyn.} \textbf{1986}, \emph{18}, 939--954\relax
\mciteBstWouldAddEndPuncttrue
\mciteSetBstMidEndSepPunct{\mcitedefaultmidpunct}
{\mcitedefaultendpunct}{\mcitedefaultseppunct}\relax
\EndOfBibitem
\bibitem[Rumble(2017)]{Rumble_CRC_2017}
Rumble,~J. \emph{CRC Handbook of Chemistry and Physics}; CRC press: Boca Raton,
  Florida, 2017\relax
\mciteBstWouldAddEndPuncttrue
\mciteSetBstMidEndSepPunct{\mcitedefaultmidpunct}
{\mcitedefaultendpunct}{\mcitedefaultseppunct}\relax
\EndOfBibitem
\bibitem[Housecroft and Jenkins(2017)Housecroft, and
  Jenkins]{Housecroft_RSCA_7_2017}
Housecroft,~C.~E.; Jenkins,~H. D.~B. Absolute Ion Hydration Enthalpies and the
  Role of Volume within Hydration Thermodynamics. \emph{RSC Adv.}
  \textbf{2017}, \emph{7}, 27881--27894\relax
\mciteBstWouldAddEndPuncttrue
\mciteSetBstMidEndSepPunct{\mcitedefaultmidpunct}
{\mcitedefaultendpunct}{\mcitedefaultseppunct}\relax
\EndOfBibitem
\bibitem[Marcus and Loewenschuss(1984)Marcus, and
  Loewenschuss]{Marcus_APRC_81_1984}
Marcus,~Y.; Loewenschuss,~A. Chapter 4. Standard Entropies of Hydration of
  Ions. \emph{Annu. Rep. Prog. Chem.{,} Sect. C: Phys. Chem.} \textbf{1984},
  \emph{81}, 81--135\relax
\mciteBstWouldAddEndPuncttrue
\mciteSetBstMidEndSepPunct{\mcitedefaultmidpunct}
{\mcitedefaultendpunct}{\mcitedefaultseppunct}\relax
\EndOfBibitem
\bibitem[Smith(1977)]{Smith_JCE_54_1977}
Smith,~D.~W. Ionic Hydration Enthalpies. \emph{J. Chem. Educ.} \textbf{1977},
  \emph{54}, 540--542\relax
\mciteBstWouldAddEndPuncttrue
\mciteSetBstMidEndSepPunct{\mcitedefaultmidpunct}
{\mcitedefaultendpunct}{\mcitedefaultseppunct}\relax
\EndOfBibitem
\bibitem[Yuan \latin{et~al.}(2017)Yuan, Ritchie, Ritter, Murphy, G{\'o}mez, and
  Mulvaney]{Yuan_JPCC_122_2017}
Yuan,~G.; Ritchie,~C.; Ritter,~M.; Murphy,~S.; G{\'o}mez,~D.~E.; Mulvaney,~P.
  The Degradation and Blinking of Single \ch{CsPbI3} Perovskite Quantum Dots.
  \emph{J. Phys. Chem. C} \textbf{2017}, \emph{122}, 13407--13415\relax
\mciteBstWouldAddEndPuncttrue
\mciteSetBstMidEndSepPunct{\mcitedefaultmidpunct}
{\mcitedefaultendpunct}{\mcitedefaultseppunct}\relax
\EndOfBibitem
\bibitem[Zhang \latin{et~al.}(2018)Zhang, Chen, Xu, Xiang, Gong, Walsh, and
  Wei]{Zhang_CPL_35_2018}
Zhang,~Y.; Chen,~S.; Xu,~P.; Xiang,~H.; Gong,~X.; Walsh,~A.; Wei,~S. Intrinsic
  Instability of the Hybrid Halide Perovskite Semiconductor \ch{CH3NH3PbI3}.
  \emph{Chin. Phys. Lett.} \textbf{2018}, \emph{35}, 036104\relax
\mciteBstWouldAddEndPuncttrue
\mciteSetBstMidEndSepPunct{\mcitedefaultmidpunct}
{\mcitedefaultendpunct}{\mcitedefaultseppunct}\relax
\EndOfBibitem
\bibitem[Hautier \latin{et~al.}(2012)Hautier, Ong, Jain, Moore, and
  Ceder]{Hautier_PRB_85_2012}
Hautier,~G.; Ong,~S.~P.; Jain,~A.; Moore,~C.~J.; Ceder,~G. Accuracy of Density
  Functional Theory in Predicting Formation Energies of Ternary Oxides from
  Binary Oxides and Its Implication on Phase Stability. \emph{Phys. Rev. B}
  \textbf{2012}, \emph{85}, 155208\relax
\mciteBstWouldAddEndPuncttrue
\mciteSetBstMidEndSepPunct{\mcitedefaultmidpunct}
{\mcitedefaultendpunct}{\mcitedefaultseppunct}\relax
\EndOfBibitem
\bibitem[Zhu and Ertekin(2019)Zhu, and Ertekin]{Zhu_EES_12_2019}
Zhu,~T.; Ertekin,~E. Mixed Phononic and Non-Phononic Transport in Hybrid Lead
  Halide Perovskites: Glass-Crystal Duality, Dynamical Disorder, and
  Anharmonicity. \emph{Energy Environ. Sci.} \textbf{2019}, \emph{12},
  216--229\relax
\mciteBstWouldAddEndPuncttrue
\mciteSetBstMidEndSepPunct{\mcitedefaultmidpunct}
{\mcitedefaultendpunct}{\mcitedefaultseppunct}\relax
\EndOfBibitem
\bibitem[Marronnier \latin{et~al.}(2018)Marronnier, Roma, Boyer-Richard,
  Pedesseau, Jancu, Bonnassieux, Katan, Stoumpos, Kanatzidis, and
  Even]{Marronnier_ACSNN12_2018}
Marronnier,~A.; Roma,~G.; Boyer-Richard,~S.; Pedesseau,~L.; Jancu,~J.-M.;
  Bonnassieux,~Y.; Katan,~C.; Stoumpos,~C.~C.; Kanatzidis,~M.~G.; Even,~J.
  Anharmonicity and Disorder in the Black Phases of Cesium Lead Iodide Used for
  Stable Inorganic Perovskite Solar Cells. \emph{ACS nano} \textbf{2018},
  \emph{12}, 3477--3486\relax
\mciteBstWouldAddEndPuncttrue
\mciteSetBstMidEndSepPunct{\mcitedefaultmidpunct}
{\mcitedefaultendpunct}{\mcitedefaultseppunct}\relax
\EndOfBibitem
\bibitem[Bonner(1981)]{Bonner_JCSFT_77_1981}
Bonner,~O.~D. Osmotic and Activity Coefficients of Methyl-Substituted Ammonium
  Chlorides. \emph{J. Chem. Soc. Faraday Trans.} \textbf{1981}, \emph{77},
  2515--2518\relax
\mciteBstWouldAddEndPuncttrue
\mciteSetBstMidEndSepPunct{\mcitedefaultmidpunct}
{\mcitedefaultendpunct}{\mcitedefaultseppunct}\relax
\EndOfBibitem
\bibitem[Belv{\`e}ze \latin{et~al.}(2004)Belv{\`e}ze, Brennecke, and
  Stadtherr]{Belveze_IECR_43_2004}
Belv{\`e}ze,~L.~S.; Brennecke,~J.~F.; Stadtherr,~M.~A. Modeling of Activity
  Coefficients of Aqueous Solutions of Quaternary Ammonium Salts with the
  Electrolyte-NRTL Equation. \emph{Ind. Eng. Chem. Res} \textbf{2004},
  \emph{43}, 815--825\relax
\mciteBstWouldAddEndPuncttrue
\mciteSetBstMidEndSepPunct{\mcitedefaultmidpunct}
{\mcitedefaultendpunct}{\mcitedefaultseppunct}\relax
\EndOfBibitem
\end{mcitethebibliography}
\bibliographystyle{manuscript}
\providecommand{\latin}[1]{#1}
\makeatletter
\providecommand{\doi}
  {\begingroup\let\do\@makeother\dospecials
  \catcode`\{=1 \catcode`\}=2 \doi@aux}
\providecommand{\doi@aux}[1]{\endgroup\texttt{#1}}
\makeatother
\providecommand*\mcitethebibliography{\thebibliography}
\csname @ifundefined\endcsname{endmcitethebibliography}
  {\let\endmcitethebibliography\endthebibliography}{}

\end{document}